\documentclass[final]{aa}
\usepackage[final]{epsfig} 

\usepackage{times}
\usepackage{latexsym}
\usepackage{amssymb}

\newcommand{\Htwo}{$\rm H_2$}
\newcommand{\tastar}{\hbox {$T_{ A}^*$ }}

\newcommand{\TMBp}{$T_{MB}$}
\newcommand{\TMB}{{$T_{MB}$} }
\newcommand{\TEX}{$T_{ex}$}

\newcommand{\cgn}{\hbox {CG~12-N }}
\newcommand{\cgnp}{\hbox {CG~12-N}}
\newcommand{\cgs}{\hbox {CG~12-S }}
\newcommand{\cgsp}{\hbox {CG~12-S}}
\newcommand{\cgsw}{\hbox {CG~12-SW }}
\newcommand{\cgswp}{\hbox {CG~12-SW}}

\newcommand{\twco}{{\hbox {\ensuremath{\mathrm{^{12}CO}} }}}
\newcommand{\twcop}{{\hbox {\ensuremath{\mathrm{^{12}CO}}}}}
\newcommand{\ceo}{{\hbox {\ensuremath{\mathrm{C^{18}O}} }}}

\newcommand{\ctfs}{{\hbox {\ensuremath{\mathrm{C^{34}S}}}}}
\newcommand{\thco}{{\hbox {\ensuremath{\mathrm{^{13}CO}} }}}

\newcommand{\htcopl}{{\hbox {\ensuremath{\mathrm{H^{13}CO^+}} }}}

\newcommand{\htcoplp}{{\hbox {\ensuremath{\mathrm{H^{13}CO^+}}}}}
\newcommand{\nthp}{{\hbox {\ensuremath{\mathrm{N_2H^+}} }}}

\newcommand{\htcopoz}{{\hbox {\ensuremath{\mathrm{H^{13}CO^+(1-0)}} }}}
\newcommand{\htcopozp}{{\hbox {\ensuremath{\mathrm{H^{13}CO^+(1-0)}}}}}
\newcommand{\dcopl}{{\hbox {\ensuremath{\mathrm{DCO^+}}}\ }}
\newcommand{\dcoplp}{{\hbox {\ensuremath{\mathrm{DCO^+}}}}}
\newcommand{\dcopto}{{\hbox {\ensuremath{\mathrm{DCO^+(2-1)}} }}}
\newcommand{\dcoptop}{{\hbox {\ensuremath{\mathrm{DCO^+(2-1)}}}}}
\newcommand{\nthpl}{{\hbox {\ensuremath{\mathrm{N_2H^+}} }}}
\newcommand{\kmps}{\ensuremath{\mathrm{km\,s^{-1}}}}
\newcommand{\Msun}{\ensuremath{\mathrm{M}_\odot}}
\newcommand{\kkms}{\ensuremath{\mathrm{K\,km\,s^{-1}}}}
\newcommand{\hour}{\ensuremath{^\mathrm{h}}}
\newcommand{\minute}{\ensuremath{^\mathrm{m}}}

\newcommand{\tautot}{\ensuremath{\tau_{\rm tot}}}
\begin{document}
   \title{The structure of the cometary   globule  CG~12: a high latitude star forming region 
\thanks{Based 
on observations collected at the European Southern Observatory,  La Silla, Chile}\thanks{Figure 
\ref{figure:c18o_1-0_channel} and
Appendix \ref{sec:PMF_app} are only available in electronic form via http://www.edpsciences.org}}

   \author{L. K. Haikala\inst{1,2}
          \and M. Olberg\inst{3}
          }

   \offprints{L. Haikala}

   \institute{Observatory, 
PO Box 14, University of Helsinki, Finland\\
              \email{haikala@astro.helsinki.fi}
         \and
Swedish-ESO Submillimetre Telescope, 
European Southern Observatory, Casilla 19001, Santiago, Chile
%         \and
%              Institute for Astronomy, University of Hawaii, 
%640 N. Aohoku Place, Hilo, HI 96720, USA\\
%              \email{reipurth@ifa.hawaii.edu}
          \and
              Onsala Space Observatory, S 439 00 Onsala, Sweden\\
              \email{michael.olberg@chalmers.se}}

   \date{Received ; accepted }

\titlerunning{The structure of the cometary   globule  CG~12}

   \abstract{ 
 The structure of the high galactic latitude Cometary Globule 12
(CG~12) has been investigated by means of radio molecular line
observations.  Detailed, high signal to noise ratio maps in
C$^{18}$O~(1--0), C$^{18}$O~(2--1) and molecules tracing high density
gas, CS (3--2), \dcopl (2--1) and \htcopl (1--0), are presented. The
C$^{18}$O line emission is distributed in a $10\arcmin$ long
North-South elongated lane with two strong maxima, \cgnp(orth) and
\cgsp(outh).  In \cgs the high density tracers delineate a compact
core, \dcopl core, which is offset by 15\arcsec \ from the \ceo maximum. 
 The observed strong \ceo emission traces the surface of the \dcopl core or a
separate, adjacent cloud component.  The driving source of the
collimated molecular outflow detected by \cite{white1993} is located
in the \dcopl core. The \ceo lines in \cgs have low intensity wings
possibly caused by the outflow.The emission in high density tracers
is weak in \cgn and especially the \htcoplp, \dcopl and \nthp lines are   
+0.5 \kmps \  offset in velocity with respect to the \ceo lines.
Evidence is presented that the molecular gas is
highly depleted. The observed strong \ceo emission towards \cgn
originates in the envelope of this depleted cloud component  or in a separate 
entity seen in the same line of sight. 
%The \ceo (1--0) to \ceo (2--1) line ratio in \cgn and in an
%area South of \cgs indicate either a very low \ceo excitation
%temperature or non LTE excitation. 
The \ceo lines in CG 12 were analyzed using Positive Matrix
Factorization, PMF.  The shape and the spatial distribution of the
individual PMF factors fitted separately to the \ceo (1--0) and (2--1)
transitions were consistent with each other.  The results indicate a
complex velocity and line excitation structure in the cloud.  Besides
separate cloud velocity components the \ceo line shapes and
intensities are influenced by excitation temperature variations caused
by e.g, the molecular outflow or by molecular depletion.  Assuming a
distance of 630 pc the size of the CG 12 compact head, 1.1 pc by 1.8
pc, and the \ceo mass larger than 100 M$_{\sun}$ are comparable to
those of other nearby low/intermediate mass star formation regions.

   \keywords{ clouds --  ISM  molecules -- ISM: structure -- 
              radio  lines -- ISM: individual objects: CG 12, NGC5367  }
   }

   \maketitle
%
%________________________________________________________________

\section{Introduction}

 Herschel~(1847) noted that a 10$^{\rm th}$ mag. star, now
 known as \object{h4636} or \object{CoD~--39\degr~8581}, is a binary
 with a separation of 3\farcs7.
% The binary is located at the edge of the
%uncertainty ellipse of a bright IRAS point source
%IRAS 13547--3944.
In the optical, h4636 illuminates the bright reflection nebula
\object{NGC~5367}. \cite{HawardenBrand1976} showed that it lies in the head of
an impressive \object{Cometary Globule 12}, \object{CG~12}, with a tail stretching about
one degree to the SE. With a galactic latitude of 21\degr \ and at the
distance of 630~pc estimated by \cite{williams1977}, CG~12 lies more
than 200~pc above the plane.  CG 12 has an associated low/intermediate
mass stellar cluster which has at least 9 members (Williams et
al. 1977). The clouds  cometary structure could be due to the passage of a
supernova blast wave. Curiously, the cometary tail stretches towards
the Galactic plane which would place the putative supernova even
farther away from the Galactic plane than the globule.  According to
\cite{maheswar2004} the head of CG~12 is pointing towards the centre
of an HI shell. Such a shell is, however, not readily
evident in the whole sky HI survey~(\cite{kalberlaetal2005}) which
merges the northern Leiden/Dwingeloo Survey~(\cite{hartmanburton1997})
and the southern Instituto Argentino de Radioastronomia
Survey~(\cite{arnaletal2000}).  The CG~12 cloud cometary shape is also
seen in the IRAS surface emission~(\cite{odenwald1988}).

   \begin{figure*} \centering \includegraphics[bb=50 10 480 700,
%   \begin{figure*} \centering \includegraphics[bb=50 100 520 700,
%     width=13.0cm,angle=-90, clip]{co_plot.ps}
     width=13.0cm,angle=-90, clip]{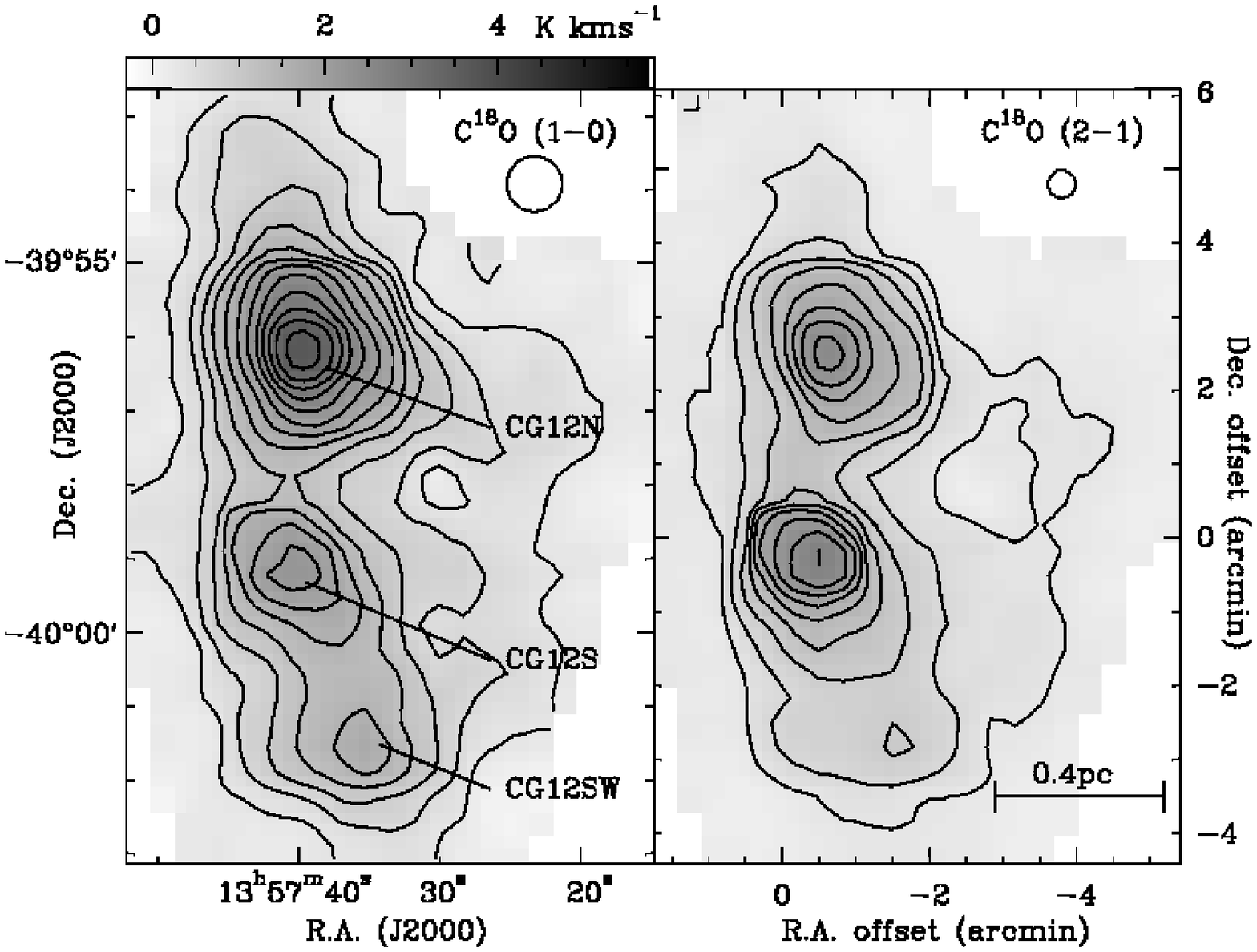}
   \caption{ Maps of integrated emission of C$^{18}$O~(1--0) and
    C$^{18}$O~(2--1) $\int$\tastar d{\it v} from $-8$ \kmps \ to $-4$ \kmps \ 
    in CG12.  The lowest contour and the contour increment are 0.3
    $\kkms$.  The SEST half power beam sizes are shown in the upper
    corners of the panels.  Offsets from the map centre position,
    $13\hour 57\minute 43\fs 1, -39\degr 58\arcmin 43\farcs 3
    $~(J2000), are shown on the axes of the right panel.  The scale
    tick in the lower right corner assumes a distance of 630 pc to the
    cloud.}
    \label{figure:CG12_CO_maps} 

\end{figure*}

 White~(1993) mapped the region around and south of h4636 in
$^{12}$CO~(2--1) and C$^{18}$O~(2--1) lines with a spatial resolution
of 22\arcsec.  He found a small size C$^{18}$O core near the binary,
and a molecular outflow with a centre close to the binary system.
Further large scale \twco, \thco and \ceo observations with 2\farcm7
resolution are presented in \cite{yonekuraetal1999a}.  CG 12 contains
two compact 1.2mm continuum sources, one in the direction of the \ceo
core detected by White~(1993) (Reipurth et al. 1996) and
another one two arcminutes north of it (Haikala 2006, in preparation).  
The centres of  the continuum sources were observed in  \ceo (3--2) 
by \cite{haikalaetal2006}.

 Near infrared~(J, H and K) images and photometry of stars in CG12
are available in  the 2MASS survey and \cite{santos1998}. A deeper 
J, H and Ks imaging study is presented in Haikala (2006, in preparation). 
Far infrared emission in CG 12 is dominated by a strong point source,
\object{IRAS 13547-3944}, near the binary h4636.

 CG 12/NGC 5367 is an intriguing object. It has the appearance of a
cometary globule, such as usually found in the outskirts of HII
regions. The linear size of cometary globules, like those in the Gum
nebula, are however much smaller (e.g., Reipurth 1983).  The linear
extent of CG12, 10 pc, is four times larger than, e.g., the archetype
object CG 1. A cluster of low/intermediate mass stellar cluster is
associated with CG 12. 
% Associated nebulosity, especially the bright
%reflection nebula NGC 5367, around some of the cluster members
%confirms the association of these stars with CG 12.  
Perhaps the most
curious feature of CG 12 is its location over 200 pc above the Galactic
plane with no sign of other nearby dark clouds or star formation.  CG
12 has not attracted much interest since the identification of the
stellar cluster (Williams et al.1977) and the \cite{HawardenBrand1976}
cometary globule paper. The subsequent papers have either concentrated
on the Herbig~AeBe binary, h4636 or the cloud has been included in
various surveys. The \twco (2--1) and \ceo (2--1) observations by
White~(1993), though detailed, cover only the very centre of the
cloud. What has been missing is a detailed but at the same time
extended study of the dense CG 12 molecular cloud in molecular
transitions sensitive to the large scale structure (cloud envelope)
and to the detailed structure of the high density material (cores).

In this paper we report the mapping of the head of CG12 in the
C$^{18}$O~(1--0) and (2--1) and in \thco (1--0) lines.
The  \ceo emission maxima  were further mapped in \dcoptop,
\htcopozp \ and CS~(2--1) and (3--2).  Further pointed
observations with long integration time in CO~(1--0) and (2--1) (and
isotopologues), CS~(2--1) and (3--2), \ctfs~(2--1), \htcopl (1--0), \dcopto
and \nthp (1--0) were made towards selected positions in the cloud.

  Observations, data reduction and calibration procedures are described in 
Sect.~\ref{sec:observations} and the observational results in 
Sect~\ref{sec:results}. The new results are compared with the optical and NIR 
images  in Sect.~\ref{sec:compare}. In 
Sect.~\ref{sec:finescale} and Appendix \ref{sec:PMF_app} Positive Matrix Factorization is used to
analyze the \ceo small scale structure in the globule. 
In the discussion part in Sect.~\ref{sec:discussion} the \ceo column densities
and the cloud mass are derived and the observational and calculated 
results are summarized. The conclusions
are drawn in  Sect.~\ref{sec:conclusions}.

%__________________________________________________________________

\section{Observations}   \label{sec:observations}

 The observations were made during various observing runs  with
the Swedish-ESO-Submillimetre-Telescope, SEST, at the La Silla
observatory, Chile. The SEST 3mm dual polarization, single
sideband~(SSB), Schotky receiver was used for the \thco (1-0) and CS
(2-1) mapping observations.  Rest of the observations were conducted
with the SEST 3 and 2~mm~(SESIS) and 3 and 1~mm~(IRAM) dual SiS SSB
receivers.  The SEST high resolution 2000 channel acousto-optical
spectrometer (bandwidth 86\,MHz, channel width 43\,kHz) was split into
two halves to measure two receivers simultaneously. At the observed
wavelengths, 3mm, 2mm and 1mm, the 43\,kHz channel width corresponds to
$\sim$0.12~\kmps, $\sim$0.08~\kmps\ and $\sim$0.06~\kmps,
respectively.

  Frequency switching observing mode was used  and a
second order baseline was subtracted from the spectra after folding.
 Calibration was achieved by the chopper wheel method. All the line
temperatures in this paper, if not specially noted, are in the
units of \tastar, i.e. corrected to outside of the atmosphere but
not for beam coupling.  Typical values for the effective SSB system
temperatures outside the atmosphere ranged from 200~K to 350~K.
Pointing was checked regularly in continuum mode towards the nearby
Centaurus A galaxy.  Pointing accuracy is estimated to be better than
5\arcsec.

The observed molecular transitions, their frequencies, SEST half power
beam width, HPBW, and the telescope main beam efficiency, $\eta_{\rm
mb}$, at these frequencies are given in Table \ref{table:obs}.  For
the CO observations only the \ceo is listed.  At a distance of 630 pc
the SEST HPBW at 219 GHz corresponds to 0.07 pc.

The cloud was mapped simultaneously in C$^{18}$O~(1--0) and
C$^{18}$O~(2--1) transitions with a spacing of 20\arcsec \ (514
positions) .  The map centre was $13\hour 57\minute 43\fs 1, -39\degr
58\arcmin 43\farcs 3 $~(J2000) which is $-5$\arcsec \ and
$\sim-$10\arcsec \ in right ascension from the positions of the IRAS
13547-3944 point source and the binary h3626, respectively. The
average rms of the spectra were 0.07~K and 0.09~K for \ceo (1--0) and
(2--1), respectively. An approximately 8\arcmin \ by 20\arcmin \ area
was mapped in \thco using 40\arcsec \ spacing (324 positions).  The
C$^{18}$O maxima were mapped in CS (2--1), (3--2), DCO$^+$(2--1) and
H$^{13}$CO$^+$(1--0).  Further long integration time pointed
observations in CO, CS, \htcoplp, \dcopl \ \ctfs \ and \nthp were
made.

%   \begin{figure*} \centering \includegraphics[bb= 17 478 320 696,
%   width=23cm,angle=270, clip]{fig2medium.ps} \caption{\tastar channel
%   map of the C$^{18}$O~(1--0) emission~(upper three rows) and
%   C$^{18}$O~(2--1)~(lower three rows).  The LSR velocity is indicated
%   in the upper left corners of the panels. Each pixel corresponds to
%   a single observed position but the C$^{18}$O~(2--1) data is binned
%   in this figure to the same channel width as the (1--0)
%   transition, 0.116 \kmps.  The highest intensity in the panels is
%   $\sim3.0$~K.  The cross in the panels is located in the map centre
%   position.}
%   \label{figure:c18o_1-0_channel} 
%   \end{figure*}

\begin{table}
%  \centering
  \caption[]{Observed lines and telescope parameters}
   \label{table:obs}
\begin{tabular}{lrcc}
\hline
\hline
Line & $\nu$ [GHz] & HPBW [$\arcsec$]& $\eta_{\rm mb}$ \\
\hline
\htcopl (1--0)   & 86.754   & 57&  0.75\\
\nthpl (1--0)     & 93.176   & 54&      \\
\ctfs \, (2--1)           & 96.412   & 54 &     \\
CS (2--1)         & 97.271   & 54&      \\
\ceo (1--0)       & 109.782  & 47&  0.70\\
\dcopl (2--1)     & 144.077  & 34&      \\
CS (3--2)         & 145.904  & 34&  0.66\\
%\dcopl (3--2)    & 216.113  & xx\\
\ceo (2--1)       & 219.560  & 24&  0.50 \\

\hline
\label{figure:table1} 
\end{tabular}                                          

\end{table}

\section{Results}\label{sec:results}  

\subsection{CO}

  The observed distributions of C$^{18}$O~(1--0) and
 C$^{18}$O~(2--1) line emission towards CG12 are presented in Fig.
 \ref{figure:CG12_CO_maps}~(the $\int$\tastar d{\it v}  in the velocity
 range $-$8 \kmps \ to $-$4 \kmps)  and
 Fig. \ref{figure:c18o_1-0_channel} ~(C$^{18}$O channel map).  In both 
 figures the grey/colour scales and the contour levels are the same
 for the two \ceo transitions.  The offsets from the map centre
 position in arc minutes are shown on the axes of the right panel in
 Fig.  \ref{figure:CG12_CO_maps}.

   \begin{figure} \centering \includegraphics  %[width=8cm,angle=-90, clip]
[width=14cm, angle=-90] {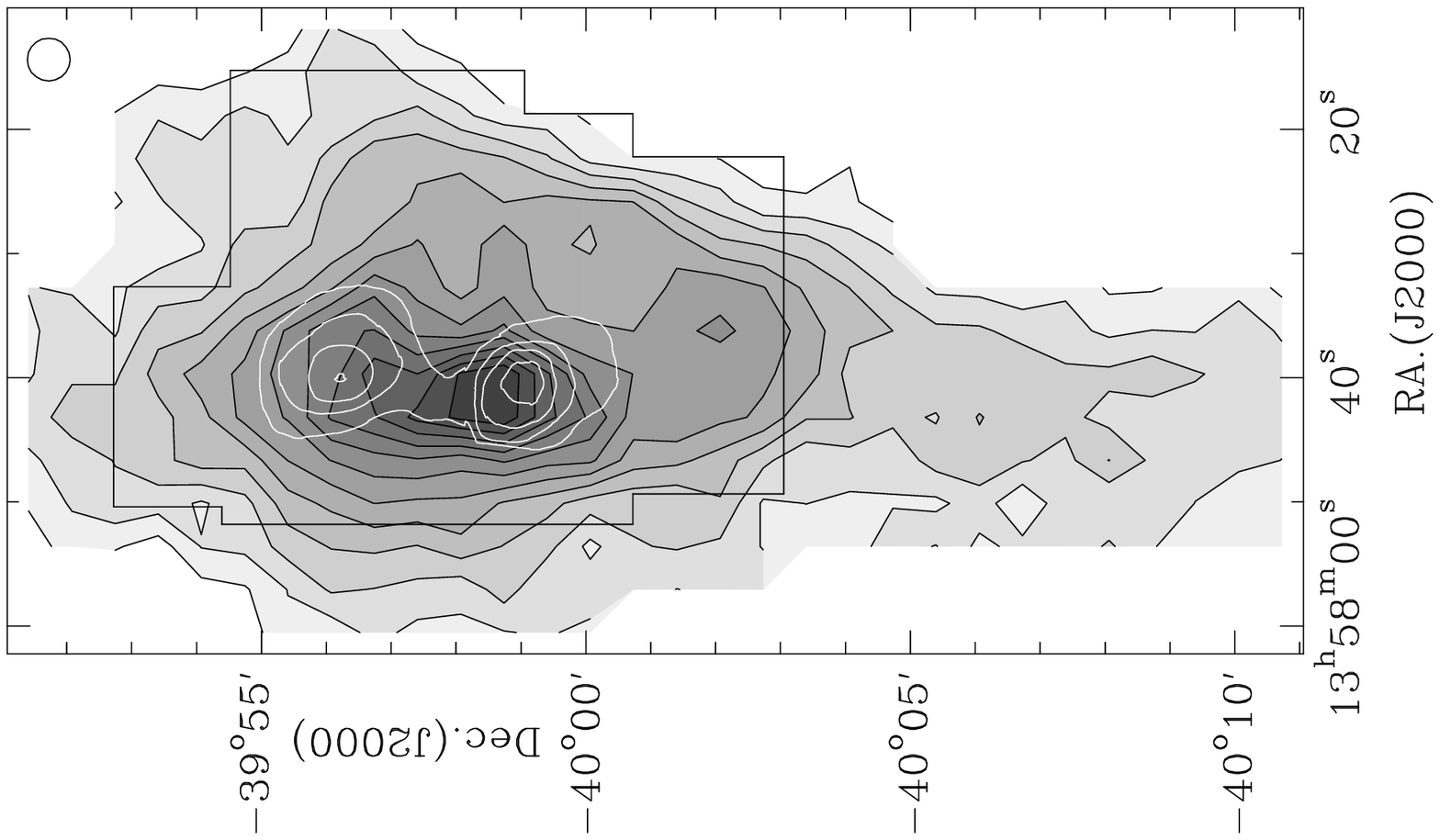}
   \caption{ Map of integrated emission of 
    $^{13}$C0~(1--0): $\int$\tastar d{\it v} from $-8$ \kmps \ to $-4$ \kmps.
    The lowest contour and the contour increment is
    1.2 $\kkms$.  The SEST half power beam size at the line
     frequency is shown in the upper corner. The extent of the \ceo mapping 
    and the outline of the \ceo ~(2-1) emission is indicated in the overlay.}
    \label{figure:CG12_13CO_map} 

\end{figure}

The bulk of the molecular material traced by C$^{18}$O emission is
distributed in a narrow North-South oriented lane with two prominent
maxima. A less intense maximum is observed to the SW.
Henceforth the three intensity maxima will be referred to as CG~12-N, 
CG~12-S and CG~12-SW. CG~12-S corresponds to the \ceo~(2--1) maximum
reported in \cite{white1993}.

   \begin{figure*} \centering \includegraphics [bb= 50 20 550 750,
width=13cm,angle=-90, clip] {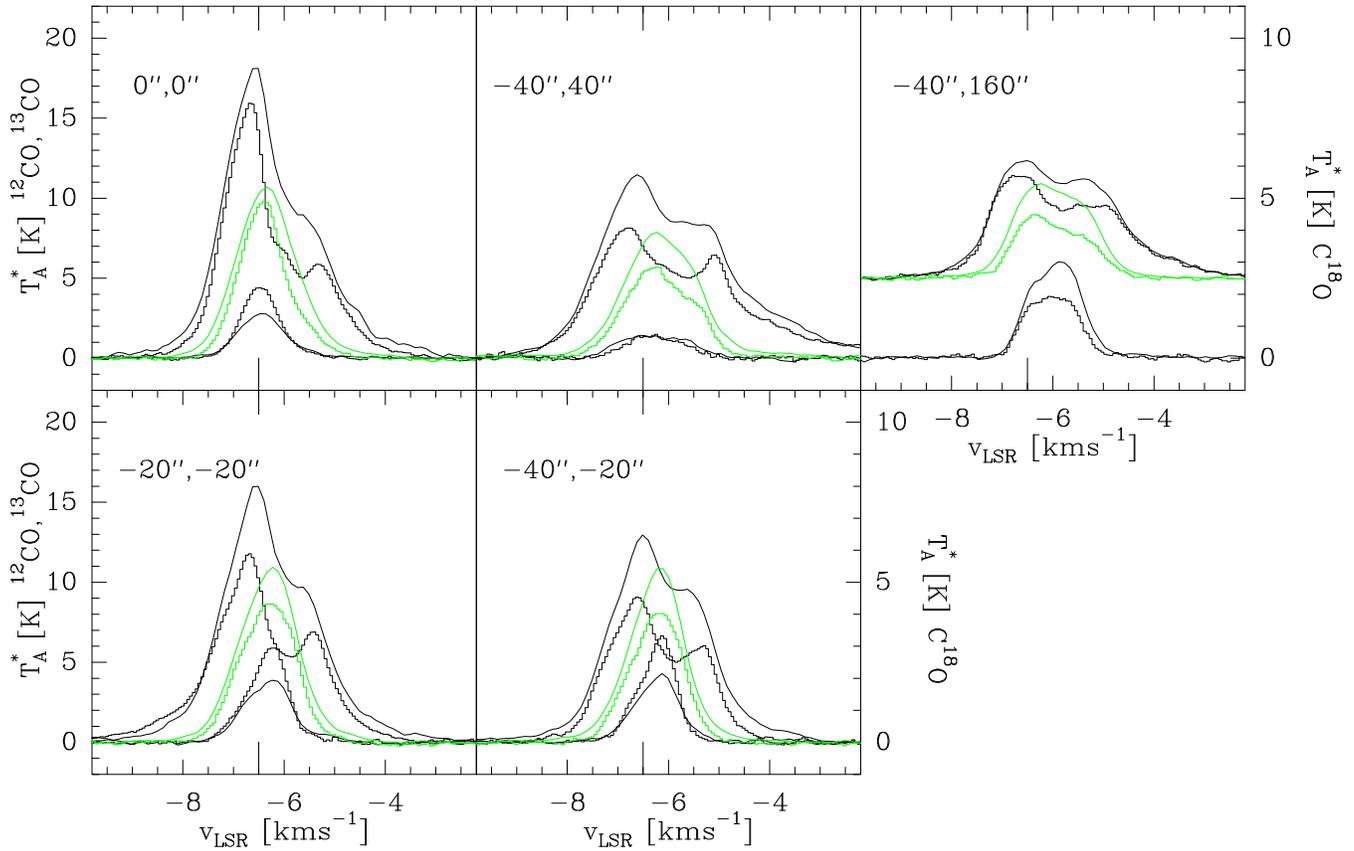}
\caption{\twco, \thco and \ceo spectra observed towards five selected
positions in CG~12. Offset from the map centre position is shown in arc
seconds in the upper left corners of each panel.  Line has been used
for the (1--0) and histogram for the (2--1) transitions.  The antenna
temperature scale at left is for $^{12}$CO and $^{13}$CO spectra and
the one on right for C$^{18}$O.  The velocity tick indicates velocity
of $-$6.5 $\kmps$.  The \twco and \thco spectra in position
($-$40\arcsec,160\arcsec) are offset from the zero level for clarity}
\label{figure:spectra}
\end{figure*}

 The morphology of the cloud is similar in the two C$^{18}$O
transitions.  However, there is a notable difference between the \ceo
maxima.  The observed C$^{18}$O~(1--0) emission is stronger than the
C$^{18}$O~(2--1) emission in \cgn and \cgsw whereas in \cgs the
opposite is the case. In \cgn the \ceo (1--0) line remains stronger
than the (2--1) line even when expressed in the main beam brightness
temperature scale.

 Further details can be seen if Fig. \ref{figure:c18o_1-0_channel}
which reveals that the structure of the cloud is not as simple as Fig.
\ref{figure:CG12_CO_maps} suggests. The most noticeable  features are
the following:  \cgn is elongated in the North South direction at
velocities $\leq -6.0$ \kmps \ and in the East West direction at
velocities $\geq -5.4$ \kmps. The position of the maximum \ceo (2--1)
emission in \cgs moves from below the map centre position (marked with
a cross in the Figure) at $-7.0$ \kmps \ to a position 40\arcsec \ west of
this position at velocity $-5.9$ \kmps.  An arc like feature connects
\cgs and \cgsw at velocities from $-6.6$ \kmps\ to $-6.2$ \kmps. The arc
is less pronounced in the \ceo (2--1) transition.

The observed distribution of the \thco ~(1--0) line emission towards
 CG12 is presented in Fig.  \ref{figure:CG12_13CO_map}. The extent of
 the \ceo mapping and the outline of the \ceo ~(2-1) emission is
 indicated in the overlay.  The notable difference between the
 observed \ceo and \thco emission is that there is only one  \thco
maximum which is offset to  NE from \cgsp. 
 In particular there is no indication of \cgn in the \thco map.

$^{12}$CO, $^{13}$CO and C$^{18}$O (1--0) and (2--1)
(pointed, long integration time) spectra  
in five selected positions in the cloud are shown in
Fig.~\ref{figure:spectra}. 
The \twco spectra observed at all positions in the figure 
are strongly self-absorbed. Therefore the \twco line
peak intensity, line half width and the line integral are not
physically meaningful.  Line wing emission due to a molecular outflow
(White 1993) is seen in all \twco spectra, also in the direction
of \cgn which was not covered by the \cite{white1993} observations.

\subsection{High density tracers, mapping  }

 In the optically thin case the critical density of the \ceo
(1--0) transition is $\sim$650 $cm^{-3}$ and approximately ten times
this value for the (2--1) transition (\cite{rohlfswilson}).  These
\ceo lines are therefore excited already at low densities and their
emission traces column density rather than number density. Therefore
the \ceo intensity maxima were also mapped in the high density tracers
(critical densities larger than $10^5 cm^{-3}$) CS~(2--1), \htcopoz
and \dcopto using 20\arcsec \ spacing. A 4 by 4 point map with the
same spacing was obtained in \cgs in the CS~(3--2) line.

\begin{figure} \centering \includegraphics 
%[bb=  51  17  547 725,
[%bb=  71  17  450 655,
width=8cm,angle=0, clip]{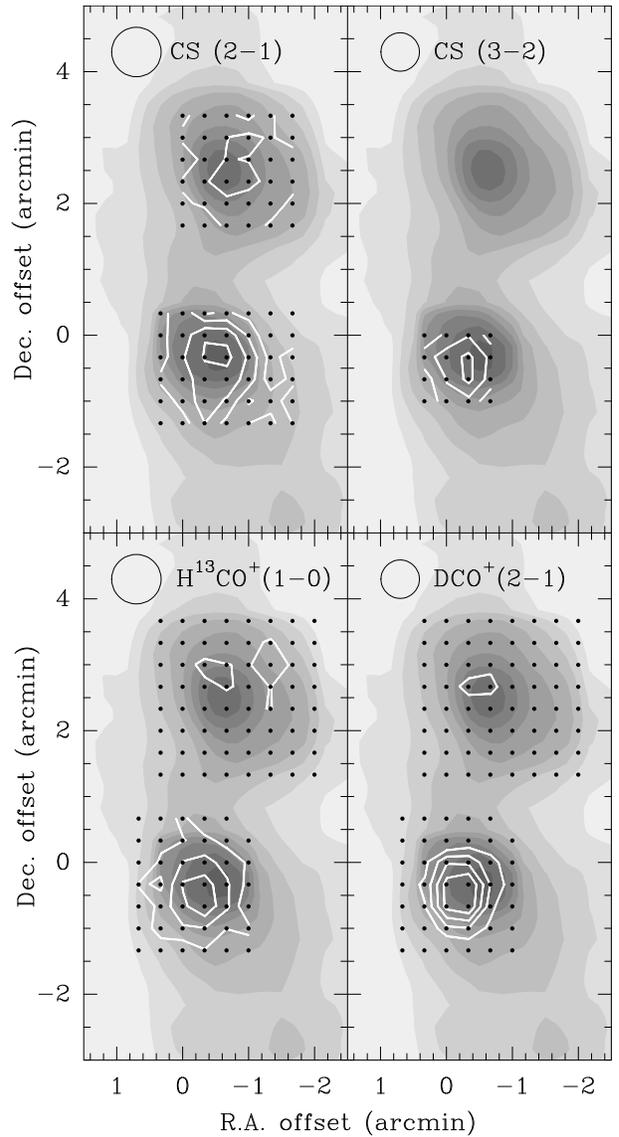}
%\begin{figure} \centering \includegraphics [bb= 90 640 400 100, 
%width=8.5cm,angle=0.0, clip]{hco_dco_ceo.ps} 
\caption{Contour maps of the observed CS~(2--1), CS~(3--2), \htcopoz
and \dcopto distributions in CG~12 superposed on a grey scale map of
the \ceo~(2--1) emission. \  The  positions observed  in each 
molecule are indicated. The corresponding SEST beam size at the
observed frequencies is indicated in the upper left corner of each
panel. In the CS~(2--1) and CS~(3--2) maps the lowest contour value is
0.5~\kkms~and the increments are 0.5 \kkms \ and 0.3 \kkms,
respectively. In the \htcopoz and \dcopto maps the lowest contour
value and its increment are 0.15~\kkms.  The maximum values of the
CS~(2--1), CS~(3--2), \htcopoz and \dcopto emission are 2.2~\kkms,
1.3~\kkms, 0.58~\kkms \ and 0.77~\kkms, respectively. The offset from
the map zero position is shown on the axes. }
\label{figure:hco_dco_ceo}
\end{figure}

 Contour maps of the CS~(2--1), CS~(3--2), H$^{13}$CO$^+$ and DCO$^+$
line integral in CG~12 superposed on the \ceo~(2--1) emission~(grey
scale) are shown in Fig. \ref{figure:hco_dco_ceo}. \cgn was not mapped
in CS~(3--2). The CS (2--1) emission peaks at the position of the \ceo
(2--1) the maximum in \cgsp. In other molecules the maximum is
shifted to SE from the \ceo maximum.  As \dcopl traces high density
gas the \cgs \dcopl maximum will be referred to as the \dcopl core.

\subsection{\ceo and high density tracers: pointed observations  } \label{hdtracers}

Pointed, long integration time  CS (2--1), (3--2), \ctfs (2--1), \htcopl
(1--0), \dcopl (2--1) and \nthp (1--0) spectra  in the same positions as in Fig
\ref{figure:spectra} are shown in Fig. \ref{figure:spectra2}. The
integration time of the CS~(3--2) line at position
($-$40\arcsec,$-$20\arcsec) is only 2.5 minutes and is therefore noisier
than the two other CS~(3--2) lines.  \nthpl (1--0) line consists of
seven hyperfine components and only the unblended \nthpl
(1--0)~(F$_1$, F$=0,1\rightarrow1,2$) component is shown in
Fig.~\ref{figure:spectra2}.

The \ctfs \ line was observed in three positions. These spectra
together with the CS~(2--1) and~(3--2) spectra observed at the same
positions are shown in Fig.~\ref{figure:spectra4}.

  The \nthpl spectra showing all the hyperfine components in the
three observed positions are shown in Fig.~\ref{figure:n2h+}. Also the
hyperfine component fits (assuming optically thin emission and 
one velocity component) and their residuals are shown.

\subsubsection{\cgs }

Near the centre of the \dcopl core~(position $-$20\arcsec,$-$20\arcsec)
the \htcoplp, \dcopl and \nthpl lines are nearly symmetric and
centered at velocity $-6.5$ \kmps.  The \ceo lines are skewed and the
maximum intensity is redshifted from the \dcopto peak velocity.  
Going to the West (position $-$40\arcsec, $-$20\arcsec) the redshifted side
of the \ceo lines becomes more intense. 
Low intensity wing-like emission is  observed in
the \ceo and  CS~(2--1) lines in the direction of the \dcopl core and 
position~(0\arcsec,0\arcsec).

The \nthpl line in the position ($-$20\arcsec,$-$20\arcsec) can be fit
with a single velocity component centred at $-$6.56 \kmps \
(Fig.~\ref{figure:n2h+}, lower panel).  Due to the 0.73 \kmps \
line half width the hyperfine components are blended and the hyperfine
structure is only marginally resolved. The signal-to-noise ratio of
the spectrum is not sufficient to make a meaningful fit of the line
total optical depth, \tautot.  The estimated \tautot \ upper limit in
this position is 5.

\subsubsection{ \cgn }

In \cgn~(position $-$40\arcsec,160\arcsec) the \ceo lines are nearly
symmetric but not gaussian.  %The C$^{18}$O~(1--0) line is seemingly
%redshifted by one channel with respect to the C$^{18}$O~(2--1) line.
The emission from the high density tracers is weak when compared to
that observed in \cgsp.

\begin{figure} \centering \includegraphics [bb= 70 50 520 600,width=8cm,
                                            angle=-90]{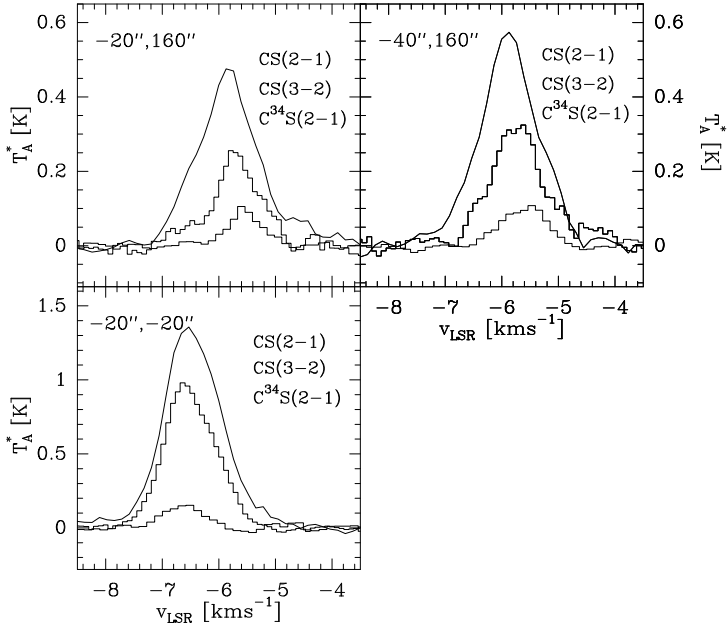}
\caption{CS~(2--1),~(3--2) and \ctfs~(1--0) spectra observed in the
direction of \cgn (the two upper panels) and \cgs (the lower panel)}.
\label{figure:spectra4}
\end{figure}

The \ctfs \ line was observed in two positions in \cgn
(Fig.~\ref{figure:spectra4}). The CS(2--1) and~(3--2) line profiles in
these positions are not gaussian and may be composed of two or three
components.  The peak intensity of the \ctfs~(2--1) line is redshifted
with respect to the peak of the CS~(3--2) line which itself is
redshifted with respect to the CS~(2--1) line. The latter line peaks
approximately at the same velocity as the \ceo lines. 

The observed ionic lines~(\dcopl (2-1), \htcopl (1-0) and \nthp
(1--0)) are offset +0.5~\kmps \ in velocity with respect to the \ceo
lines and peak at a velocity where especially the C$^{18}$O~(2--1)
line intensity is low.

\begin{figure} \centering \includegraphics [bb= 10 30 500 750,width=10cm]{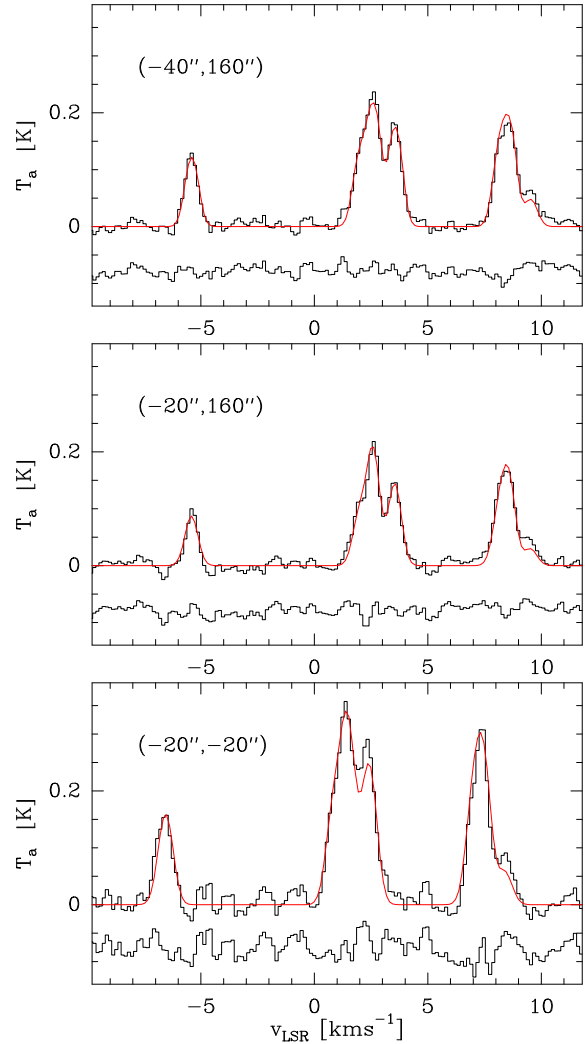}
\caption{\nthp~(1--0) spectra observed in the direction of \cgn 
(the two upper panels) and \cgs (the lower panel).  The seven
component hyperfine structure fit and the resulting residuals are
shown} \label{figure:n2h+}
\end{figure}

N$_2$H$^+$ (1--0) was observed in the same two positions as
\ctfs. Contrary to the CS lines, the observed \nthpl lines are
identical within the noise (Fig. \ref{figure:n2h+}, two upper
panels). The SEST beam is the same at CS (2--1) and \nthpl (1--0)
frequencies.  A hyperfine component fit to the \cgn \nthpl lines gives
a 0.65 \kmps \ broad line at $-5.41$ \kmps. The estimated \tautot \
upper limit in position ($-$40\arcsec,$-$160\arcsec) is 2.

 To rule out that the observed  intensity/velocity structure described above is
due to calibration, pointing or frequency setting problems, the
observations were carefully checked by observing the C$^{18}$O and the
DCO$^+$, H$^{13}$CO$^+$ and CS lines after each other and making
pointing checks before and after observations. Also, e.g., the line pair
DCO$^+$~(2--1) and \ceo (1-0) could be observed simultaneously 
with a dual receiver, thus excluding relative pointing errors.

\begin{figure*} \centering \includegraphics [ bb= 40 20 550 780,
%   \begin{figure*} \centering \includegraphics [bb= 30 20 550 750,
width=12.5cm,angle=-90, clip] {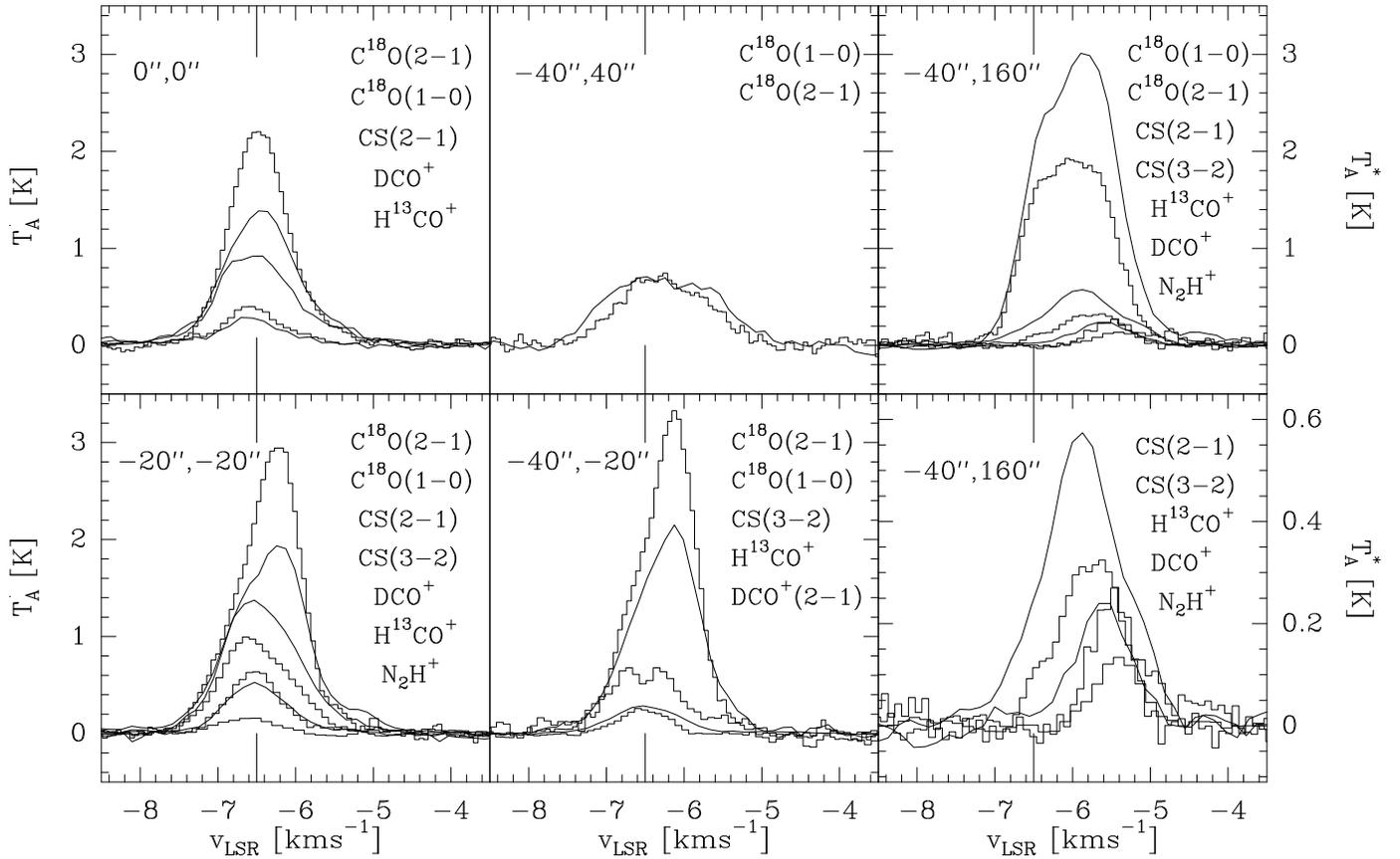}
\caption{Spectra of \ceo and molecules sensitive to number density
towards the same positions as shown in Fig. \ref{figure:spectra}.  Only
the unblended hyperfine \nthpl~(1--0) (F$_1$, F$=1,0\rightarrow1,2$)
component is shown.  The velocity tick indicates a velocity of $-$6.5
$\kmps$. \ceo (1--0), CS~(2--1) and \htcopl lines are plotted using
continuous line.  The panel lower right is a blowout of the upper
right panel showing only the lines tracing high number density.
}  \label{figure:spectra2}
\end{figure*}

%\begin{figure} \centering \includegraphics [bb= 10 50 550 600,width=6cm]{n2h+.eps}

\section{Comparison with observations at other wavelengths}\label{sec:compare}

\begin{figure} \centering \includegraphics 
[width=8cm, angle=-90]{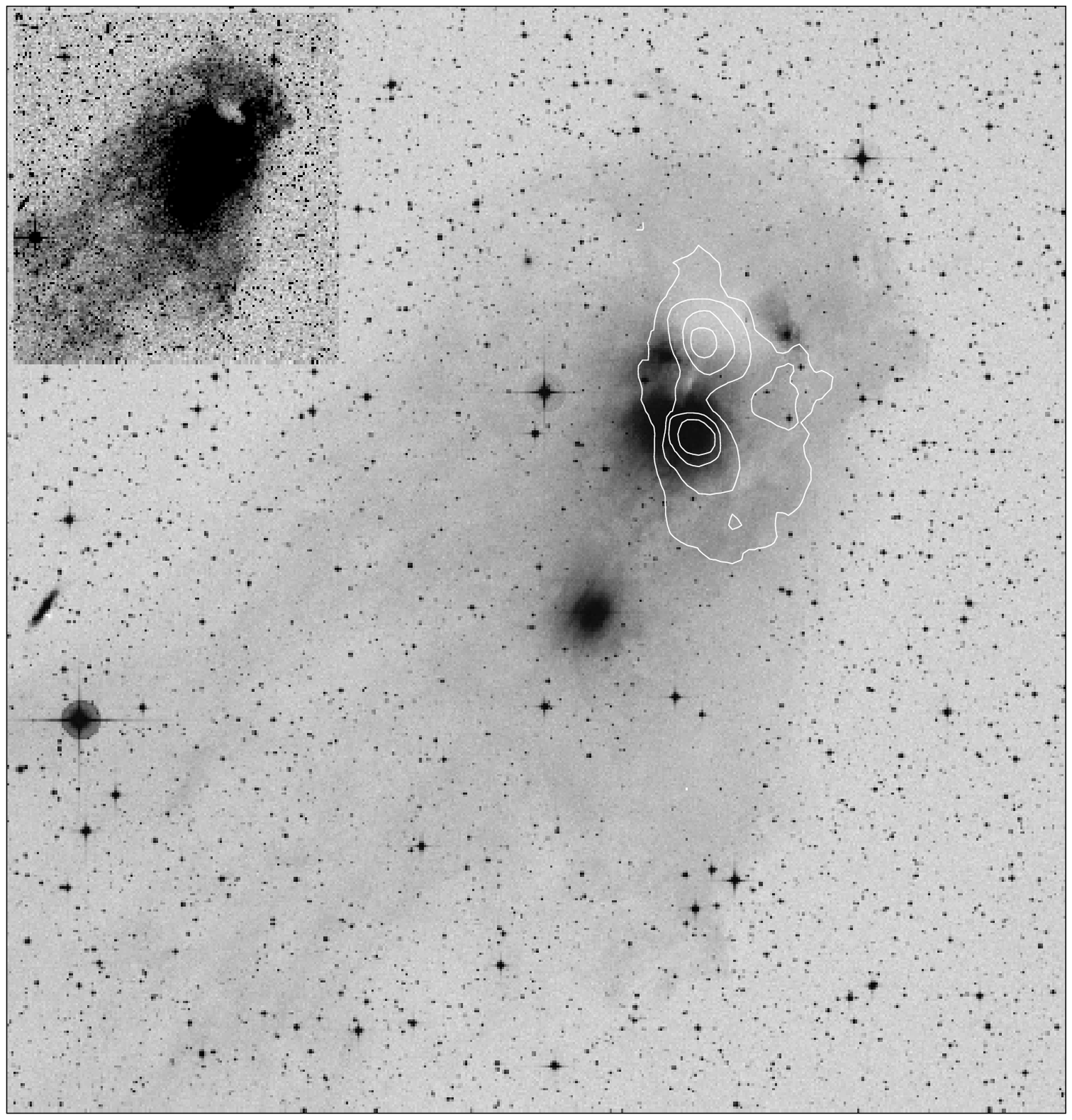}
\caption{Contour map of the \ceo~(2--1) emission superposed on the
optical~(DSS blue) image of NGC 5367. The image is also shown in the
inset but using an intensity scale which better brings out the
surface emission. The linear size of the image is 32\arcmin by
32\arcmin.}
\label{figure:cg_cerc}
\end{figure}

The \ceo~(2--1) contour map superposed on the blue  SERC-J DSS image is shown in
Fig.~\ref{figure:cg_cerc}.  The bright optical reflection nebula
NGC 5367 lies in front of \cgs whereas  \cgn is in the direction of 
an optically heavily obscured region which is clearly seen in the inset.
A 1.2 mm source (Haikala 2006, in preparation) coincides with the position of
the \cgn \ceo maximum.

\begin{figure} \centering \includegraphics 
[width=7.5cm, angle=-90]{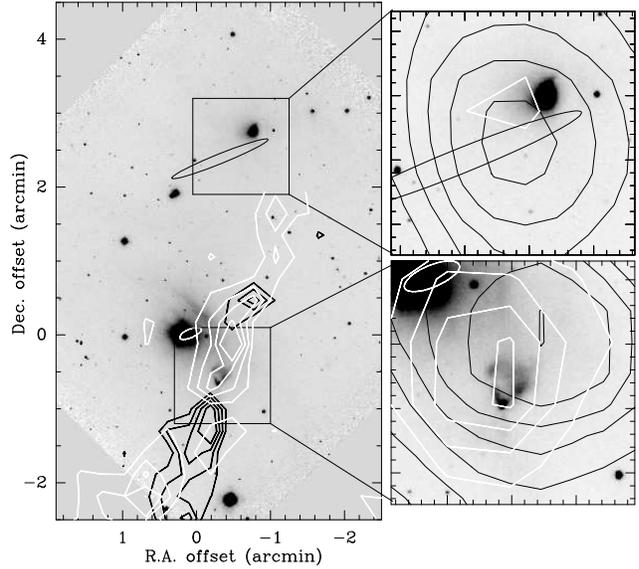}
\caption{ Ks--band image of CG-12~(Haikala 2006, in preparation).  Contours of
the blue~(white contours) and redshifted~(black) \twco~(2--1)
line wing emission (White 1993) are also shown. The two
insets show the \cgs and \cgn regions in detail.  Superposed are
the contours of \ceo~(2--1)~(black) and \dcopl ~(white)
emission. The positional uncertainty ellipses of the IRAS point sources
IRAS 13547-3944 in \cgs and IRAS 13546-3941 in \cgn are also shown.}
 \label{figure:sofi}
\end{figure}

A Ks band image~(Haikala 2006, in preparation) of CG~12 is shown in
Fig.~\ref{figure:sofi}. Superposed on the image are the contours of
the \twco (2--1)  molecular outflow
line wing emission~(White 1993).   In the insets the \ceo~(2--1) and
\dcopl contour maps are superposed together with the positional
uncertainty ellipses of IRAS point sources IRAS \object{13546-3941} and
IRAS 13547-3944.  IRAS 13547-3944 does not coincide with the
binary h4636 (the base of the binary star
emission is saturated due to the intensity scale which was chosen to
emphasize the low intensity surface emission).

A cone like nebulosity with a bright head is seen projected on the
\dcopl core in Fig.  \ref{figure:sofi}.  The apex of the cone is
located just off the tip of the red shifted and below the end of the
blue shifted, collimated CO outflow lobes of White (1993).  A compact
1.2 mm source (Haikala 2006, in preparation) coincides with the \dcopl
core. The core does not have an associated point source but this could
be due to the low spatial resolution of the IRAS satellite. A faint
source would be masked by the strong nearby source IRAS
13547-3944. The NIR cone  could be associated
with the driving source of the outflow. The centre of the molecular
outflow is offset from the position of IRAS 13547-3944 by
$\sim$20\arcsec. It is unlikely that the point source is associated
with the large outflow but it could be the driving force of the strong
redshifted outflow lobe located $\sim$1\arcmin \ NW of the point
source nominal position.
  
 Small areas of the size of the SEST HPBW at the \ceo (1--0) frequency
in the centres of \cgn and \cgs have been mapped in the \ceo (3--2)
(Haikala et al. 2006).  The HPBW of the \ceo (3--2) observations
is 19\arcsec \ which is similar to the SEST HPBW of 24\arcsec \  at the
\ceo (2--1) frequency. In \cgs the \ceo (3--2) emission has a strong
maximum in the same position as the maxima in the lower \ceo
transitions. The line profile in position ($-$20\arcsec,$-$20\arcsec)
agrees with the \ceo (2--1) line (Fig.  \ref{figure:spectra2}).  The
\ceo (3--2) \TMB peak line temperature in this position is, however,
10~K which is nearly twice the \ceo (2--1) \TMB of 5.7~K.  In \cgn
(position ($-$40\arcsec,160\arcsec) the \ceo (3--2) line profile is
similar to that of \ceo (1--0) but the \TMB peak line temperature is
3~K which is lower than the corresponding values observed in \ceo
(1--0) and (2--1) which are 4.3~K and 3.8~K, respectively.

\section{\ceo fine scale structure}\label{sec:finescale}

 The half widths of the observed C$^{18}$O lines in CG~12 cores range
from 1.0~\kmps \ (\cgsp) to 1.3 \kmps \ (\cgnp).  The \ceo
channel-maps (Fig.~\ref{figure:c18o_1-0_channel}) and observations of
other molecules indicate, that the lines consist of more than one
velocity component. The division of the cloud according to its
appearance in the \ceo line integral
maps~(Fig.~\ref{figure:CG12_CO_maps}) into only three components,
\cgnp, \cgs and \cgswp, is therefore too coarse.  The fine structure
within the individual maxima is not seen because the lines are heavily
blended in velocity.

\subsection{Positive Matrix Factorization}\label{sec:PMF}

 Positive Matrix Factorization (PMF) has been  used by \cite{juvela1996} and 
\cite{Russeiletal2003} in
the analysis of molecular line spectral maps. PMF assumes, that the ensemble
of input spectra are composed of {\it n} individual line
components~(factors).  Unlike the Principal Component Analysis PMF
assumes that the individual components are positive. This makes the
interpretation of the results more straightforward than in Principal
Component Analysis where the results may contain negative components.
No other assumptions are made of the shape of the factors.  Each input
spectrum can be reconstructed from the {\it n} PMF output factors by
multiplying each factor by the weight assigned to it by PMF at the
position and adding up the multiplied factors. 
A comparison of
the results obtained with PMF analysis with those obtained using the 
analysis of channel maps and by the use of Principal component analysis
is given \cite{Russeiletal2003}.

PMF has been applied separately to the north and south parts of
CG~12 but there is a small positional overlap between the two areas used
in the analysis.  The same dataset, which is shown in
Fig.~\ref{figure:c18o_1-0_channel} was used in the analysis, i.e, the
\ceo~(2--1) data is binned to the same channel width
as the \ceo~(1--0) data. The PMF analysis is presented in 
Appendix \ref{sec:PMF_app}.

\section{Discussion }\label{sec:discussion} 

\subsection{\ceo as a tracer of molecular material}\label{sec:c18o_interp}
 
C$^{18}$O emission is generally considered a good tracer of large
scale structure of dark clouds and globules. However, \ceo
photodissociation in the cloud envelopes and molecular depletion in
the dense and cold cloud cores may restrict the number density
interval where \ceo emission can be used as a direct measure of \Htwo \
column density (especially if the LTE approximation is used).
%The \ceo molecules are subject to photodissociation in the 
%cloud low density envelope. 
Unlike \twcop, the \ceo molecules are not shielded against
photodissociation by strong \Htwo \ lines and further, the \ceo
self-shielding is weaker than that of \twcop.
%(\cite{warinetal1996}). 
For \ceo the self-shielding is most efficient for the two lowest rotational levels
with the largest populations. According to \cite{warinetal1996}, the
consequence is that the J=1--0 transition becomes thermalized, whereas
the higher transitions remain subthermally excited in the cloud envelopes.
In the other extreme (pre-stellar cloud cores and protostellar
envelopes) the CO molecule may  vanish from gas phase
because of depletion onto dust grains. Further complications in
interpreting \ceo data are the possible strong gradients in the CO
excitation temperature on the line of sight due to heating of the gas
by newly born stars and protostellar objects. Time and temperature
dependent chemistry like, e.g., deuterium fractionation, also
complicates the comparison of \ceo data with that of other molecular
species.

 Depletion of molecules on dust grains takes place in the dense,
quiescent and cold molecular cloud cores. CO and CS are among the
first molecules to disappear from the gas phase but nitrogen bearing
species like, e.g., \nthp and ammonia, seem to be able to resist
depletion (\cite{tafallaetal2002}).  The \nthp and \dcopl lines in
\cgn peak at a velocity of 5.4~$\kmps$ \ where the \ceo line emission
is weak (Sect. \ref{hdtracers}). It is likely, that strong CO
depletion has taken place in the gas traced by these two molecules in
\cgn.  

In \cgs the \ceo maximum is spatially offset from the \dcopl core.
Deuterium fractionation reactions are favored in cold gas (e.g.,
\cite{herbst1982}) and therefore the relative abundance of \dcopl is
enhanced in cold cloud cores. The relatively strong \dcopl emission
observed in \cgsp, the \dcopl core, could indicate that fractionation
has indeed taken place.  If the molecules in \cgs were heavily
depleted in the cold \dcopl core one would not expect CS~(3--2) or
\htcopl to outline the core like they do (also \dcopl will be finally
depleted).
% It is, however, possible that a fraction of the CO
%molecules are indeed depleted in \cgsp.

 The observed \twco (1--0) and (2--1) line temperatures in \cgs
(Fig. \ref{figure:spectra}) are high. The lines are self reversed so
the actual line peak temperatures must be even higher.  This
indicates that the CO excitation temperature is in excess
of 30~K in the part of the cloud which is traced by the observed \twco
emission. 
%The \thco intensity has a maximum NE from \cgsp (Fig. \ref{figure:CG12_13CO_map}).  
White (1993) suggests that either h4636 or IRAS 13547-3944 is the
heating source.  The \twco and \thco optical depth is high and unlike
\ceo these CO isotopoloques are likely to trace only the surface of
the molecular cloud associated with CG 12.  The observed \ceo (3--2) \TMB \
peak temperature of 11 K in \cgs also points at a high \ceo excitation
temperature (Haikala et al. 2006).  However, the observed relative
\ceo (2--1) and (1--0) line intensities in \cgs are not compatible
with \ceo \TEX \ values higher than  20~K.

\subsection{PMF: The interpretation}\label{sec:pmf_interp}

One should be cautious in interpreting the PMF results. Even though
the fit to the \ceo (1--0) and (2--1) data is good, all the PMF factors do
not necessarily describe real cloud components. 
The structure of the cloud 
can, however, be discussed with some confidence when the PMF fit results are
considered with the information provided by other 
available molecular line, mm continuum  and NIR-FIR data. 

%The size, 
%40\arcsec\  by 40\arcsec,  and the signal-to-noise
%ratio of the  APEX  \ceo (3--2) maps in  \cite{haikalaetal2006}
%are not sufficient for a detailed 
%comparison. Only the map centre positions were integrated long enough
%to achieve a good signal-to-noise ratio. The spectra in these positions 
%provide an independent check on the PMF analysis.

\begin{figure} \centering \includegraphics 
[bb= 24 24 835 460, height=8.5cm]{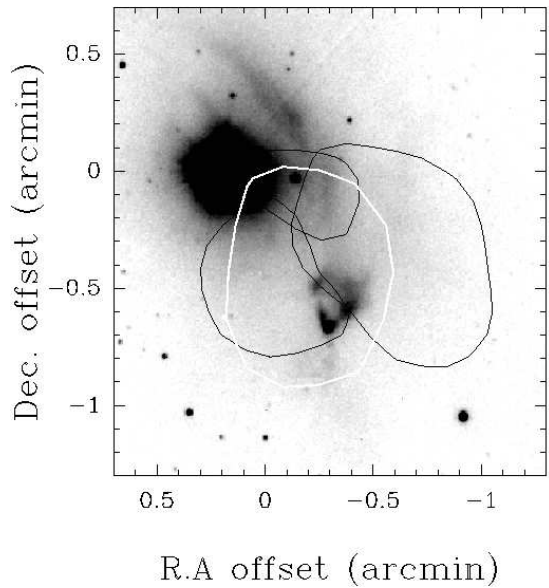}
%[bb= 54 64 835 460, width=10.5cm]{haikala_fig_10.eps}
\caption{The \cgs \ceo (2--1) 1{\it s}, 2{\it s} and 3{\it s} (from
  East to West) factor peak emission positions (0.8~K \kmps \
  contours) and \dcopl core position ~(white) superposed on the SOFI
  Ks band image. Offsets from the map zero position are shown in
  arcminutes. The SEST beam at the \ceo~(2--1) frequency is 0.37
  arcminutes}
\label{figure:sofi_factors}
\end{figure}

\subsubsection{\cgs}

 The location of the \cgs \ceo (2--1) PMF factor maxima relative to the NIR
cone (and the mm continuum source) in the centre of the \dcopl core
is shown in Fig.~\ref{figure:sofi_factors}.  
From East to West the factors are 1{\it s},
2{\it s} and 3{\it s}. The centres of the maxima differ in position by
about one SEST beam size at 220 GHz, 24\arcsec. 

The data would seem to implicate that \cgs is fragmented into three
individual cores. If this were the case one would expect that the
2{\it s} factor would coincide with the \dcopl core because the 2{\it
s} centre of line  velocity is the same as that of \dcoplp. Even though 
the 3{\it s} factor is  the strongest in \cgs it is detected  only as an
asymmetry in the \dcopl and CS lines in Figs. \ref{figure:spectra2} and
\ref{figure:spectra4}. 
%  PMF not being able
%to find a counterpart for the \dcopl core is, however, not a
%shortcoming of the method but rather that of the input data. 
%The \dcopl core is offset from the \ceo maximum
%(Fig. \ref{figure:hco_dco_ceo}). 
CO is ubiquitous and easily excited
and therefore it can be detected at much lower densities than the high
density tracers, \dcopl and CS (3--2), which trace the \dcopl core.
The bulk of the observed \ceo emission may have its origin in the part
of the cloud where the density is lower and the excitation temperature
higher than in the \dcopl core.

This leads to a model where most of the observed 
\ceo emission traces rather
the surface of the dense and  cold cloud core (the \dcopl core) or
a separate, adjacent cloud component, than the
core itself. The observed CS~(2--1) distribution, which is similar to
\ceo in \cgs (Fig.~\ref{figure:hco_dco_ceo}), is in accord with this
model.  Much of the observed CS~(2--1) emission is known to originate
in cloud envelopes and not in the dense cores. 
The CS~(3--2) effective critical density is higher
than that of CS~(2--1) and therefore it traces the dense gas deeper in
the cloud than CS~(2--1).

The possible interaction of the collimated molecular outflow detected
by \cite{white1993} with the cloud core further complicates the
interpretation of the molecular line data.  The centre of the outflow
is located in the direction of the \dcopl core and
%and the NIR cone lies projected on the tip of the blue shifted outflow lobe
%(Fig. \ref{figure:sofi}). 
it is not unlikely that the outflow can produce weak \ceo line wings
and that it can raise the CO excitation temperature locally. Weak line
wings are also observed in the CS~(2--1) line in
Fig.~\ref{figure:spectra2}.  It is argued that the weak \ceo red and
blue shifted line wing emission near the \dcopl core is produced by
the interaction of the collimated molecular outflow with the parent
cloud. The line wings would be described by the PMF factors 1{\it s}
and 4{\it s}. For the latter factor this would consists only of the
emission immediately to the west of the \dcopl core.

The \cgs \ceo (3--2) map in Haikala et al. (2006) covers only the
very centre of \cgsp.  The \ceo (3--2) line peak velocity and the
intensity distribution is in accordance with the elongated shape of
the 3{\it s} component in Fig. \ref{figure:sofi_factors}. 
%A hot, small
%size object would explain the observed \ceo (3--2) and (2--1) line
%intensities but not the \ceo (1--0) line intensity which is too strong
%(Haikala et al. 2006). 
The maximum of the 3{\it s} factor is compact
and lies on the outflow axis only $\sim$20\arcsec \ away from the apex
of the NIR cone (Fig. \ref{figure:sofi}), the putative position of the
outflow driving source. This suggest, that the high 3{\it s} \ceo line
intensities and the molecular outflow could be connected.

\subsubsection{\cgn}

PMF produces a seemingly straightforward solution for \cgnp. However,
the available data from molecules other than \ceo and the 1.2~mm
continuum emission deny such a simple solution. As argued in Sect. 
\ref{sec:c18o_interp} it is likely that the cloud component traced 
by \dcopl and \nthpl is heavily depleted.  If
this is the case \ceo only probes the undepleted part of \cgnp.

%J\o rgensen et al. (2004) report  \ceo line ratios (including the \ceo (3--2)
%line)
%similar to those in \cgn in the direction of Class 0 sources. They
%propose a model where the gas is depleted in a cold layer between a
%warm cloud surface and the centre of the core which is warmed up by a
%central source. The size of \cgn is however much larger than that of a
%circumstellar shell. 

\subsection{Cloud physical properties}\label{sec:properties}

A straightforward division of the CG 12 molecular cloud into three
homogeneous components, \cgnp, \cgs and \cgswp, is not possible. The good
velocity resolution and the high signal to noise ratio of the data
allows one to divide the \ceo data into separate components.  However,
the derivation of the cloud or core physical properties calls for a
detailed three dimensional non-LTE model which takes into account both
the molecular depletion, the varying density and excitation
conditions. Such a model is beyond the scope of this paper and will be
left for the future. The LTE-approximation approach  is used instead of
a sophisticated model to make a zeroth order estimate of the masses of
the three \ceo maxima, \cgnp, \cgs and \cgswp. An average
\ceo excitation temperature for each maximum is estimated using 
the observed relative \ceo (2--1) and (1--0) line intensities.

 \subsubsection {Mass and column density estimation} 

The observed \tastar line temperatures were converted into main beam
temperatures  using the  beam efficiencies in Table
\ref{figure:table1}.  The \ceo excitation temperatures were estimated
from the \ceo (1--0) and (2--1) data
assuming optically thin \ceo emission and LTE.  This assumes that the
observed emission in both transitions originates in the same volume of
gas at a constant excitation temperature.  The observed \ceo line
\TMBp(2--1)/ \TMBp(1--0)  ratio is 1.8 in \cgsp, compatible
with an excitation temperature near 15~K. In \cgn and \cgsw this ratio
is one or less indicating excitation temperatures of the order of 10~K.

The masses of \cgnp, \cgs and \cgsw are calculated without dividing
them into smaller components. 
The spectra between declination offsets (including
the limits) $-$120\arcsec \ and +60\arcsec \ were assigned to
\cgsp. The spectra above and below these limits were assigned to \cgn
and \cgswp, respectively.
An average \ceo excitation temperature of 10~K is now assumed for \cgn
and \cgsw and 15~K for \cgsp. The adopted \ceo abundance is 2.0
$10^{-7}$.  This results in calculated total masses of \  34, 96, and
110 \Msun \ for \cgswp, \cgs and \cgnp, respectively, when \ceo (1--0)
data is used. If \ceo (2--1) data is used the masses are 14, 49
and 53 \Msun. 
 The maximum \TMB \ line integrals  can be used to estimate the \Htwo \ column
densities in the corresponding beams.  The calculated maximum column
densities in the \ceo (1--0) beam are 1.3 $10^{22}$, 1.8 $10^{22}$ and
2.7 $10^{22}$ $cm^{-2}$ for \cgswp, \cgs and \cgnp, respectively. For
the \ceo (2--1) beam the corresponding values are 6.4 $10^{21}$, 1.6
$10^{22}$ and 1.8 $10^{22}$ $cm^{-2}$.

 The masses calculated from the \ceo (1--0) data are approximately
twice higher than from the (2--1) data. The LTE approximation assumes
that all the \ceo rotational states are thermalized. However,
according to Warin et al. (1996) only the lowest \ceo rotational
state is thermalized in dark clouds and the higher states are
subthermally excited. The subthermal excitation would be stronger in
the relatively low density cloud envelope than in the more dense
core. The subthermal level population leads
to an  underestimation of the cloud mass when calculated from the \ceo
(2--1) data.  This could be a partial explanation for the large
discrepancy between the masses calculated using \ceo (1--0) and (2--1)
data.
 
Strong \ceo (3--2) emission (maximum \TMB $\sim$11~K) distributed
similar to the 3{\it s} PMF factor was detected by
\cite{haikalaetal2006} in the centre of \cgs.
%The signal to noise
%ration of these obsevations allow only to resolve the 2{\it s} and
%3{\it s} PMF factors and the high line temperature corresponds to the
%latter factor.  
The \ceo (3--2) data was modelled with a compact (60\arcsec \ to
80\arcsec \ diameter) and hot (80~K $\lesssim $\TEX $\lesssim$ 200~K)
optically thin clump of $\sim$ 1.6 \Msun. The high temperature was
derived from the \TMB(\ceo(3--2))/\TMB(\ceo (2--1)) ratio. However,
\TMB(\ceo(2--1))/\TMB(\ceo (1--0)) ratio for the \cgs PMF factor 3{\it
s} is not compatible with such a high temperature. This could be due
to, e.g., contribution from a subthermally excited cool cloud envelope
to the \ceo (1--0) emission. The LTE mass calculated from the \ceo
(2--1) 3{\it s} component within the area modelled in
\cite{haikalaetal2006} is 4.1 \Msun \ when a \TEX \ of 15~K is
assumed.  The discrepancy in the calculated excitation temperatures
and masses highlights the uncertainties when LTE approximation is used
and only data of the two lowest \ceo transitions are available. In
contrast to \cgs the \ceo \TEX \ and the LTE mass of \cgn correspond
well to the modelled values using the the three transitions (Haikala
et al 2006).

\subsection{Summary}\label{sec:summary} 

 The analysis of the spectral lines  presented above 
deliver a complicated picture of CG 12. Even though the
spectral lines are relatively narrow the analysis
reveals a  rich structure in line shape and  velocity.
Probable depletion of molecules on dust grains in \cgn
further complicates the interpretation.  
The cloud parameters
derived using the LTE analysis are highly uncertain. However, if the 
molecular line observations are combined with the available
optical, NIR, FIR and mm continuum data the cloud structure
can be discussed with some confidence.

 \subsubsection{\cgn}

 CG 12-N harbors a compact, cold mm source and the detected relatively
weak molecular emission in high density tracers is probably associated
with this source. Much of the molecular material associated with this
cloud component is likely to be highly depleted onto dust.  The
observed strong \ceo emission towards \cgn originates therefore in the
envelope of this depleted core or in a separate entity seen in the same
line of sight.  \twco
observed in the direction of \cgn  has line wing emission
indicating molecular outflow. It is not known if this outflow is
connected to the outflow in \cgs or if it is local and originates in
\cgn.

 \subsubsection{\cgs} 
 
The bulk of the observed \ceo emission does not trace the gas
associated with the compact cloud core detected in \htcoplp, \dcopl
and CS (3--2) lines.  Most likely the \ceo emission
traces only the surface of this
core. The moderate \ceo line wing emission, probably due to
interaction of the highly collimated molecular outflow with the
surrounding gas, further complicates the interpretation.

The  molecular line data presented in this paper combined
with the molecular outflow data from \cite{white1993}, the NIR imaging
and mm dust continuum data (Haikala 2006, in preparation) shows
that the outflow centre coincides with the mm continuum source and the
NIR cone in the centre of the \dcopl core.  
The strong point source IRAS 13547--3944 is offset from the
mm source and from h4636.  This point source could, however, be the
driving source for the strong redshifted outflow lobe north of \cgsp.

 \subsubsection{\cgsw}

\begin{figure*} \centering \includegraphics
%[bb= 17 478 320 696,    width=23cm,angle=270, clip]{haikala_fig_11.eps} \caption{\tastar channel
%[bb= 110 408 550 100, width=23cm,angle=270, clip]
[height=22cm]{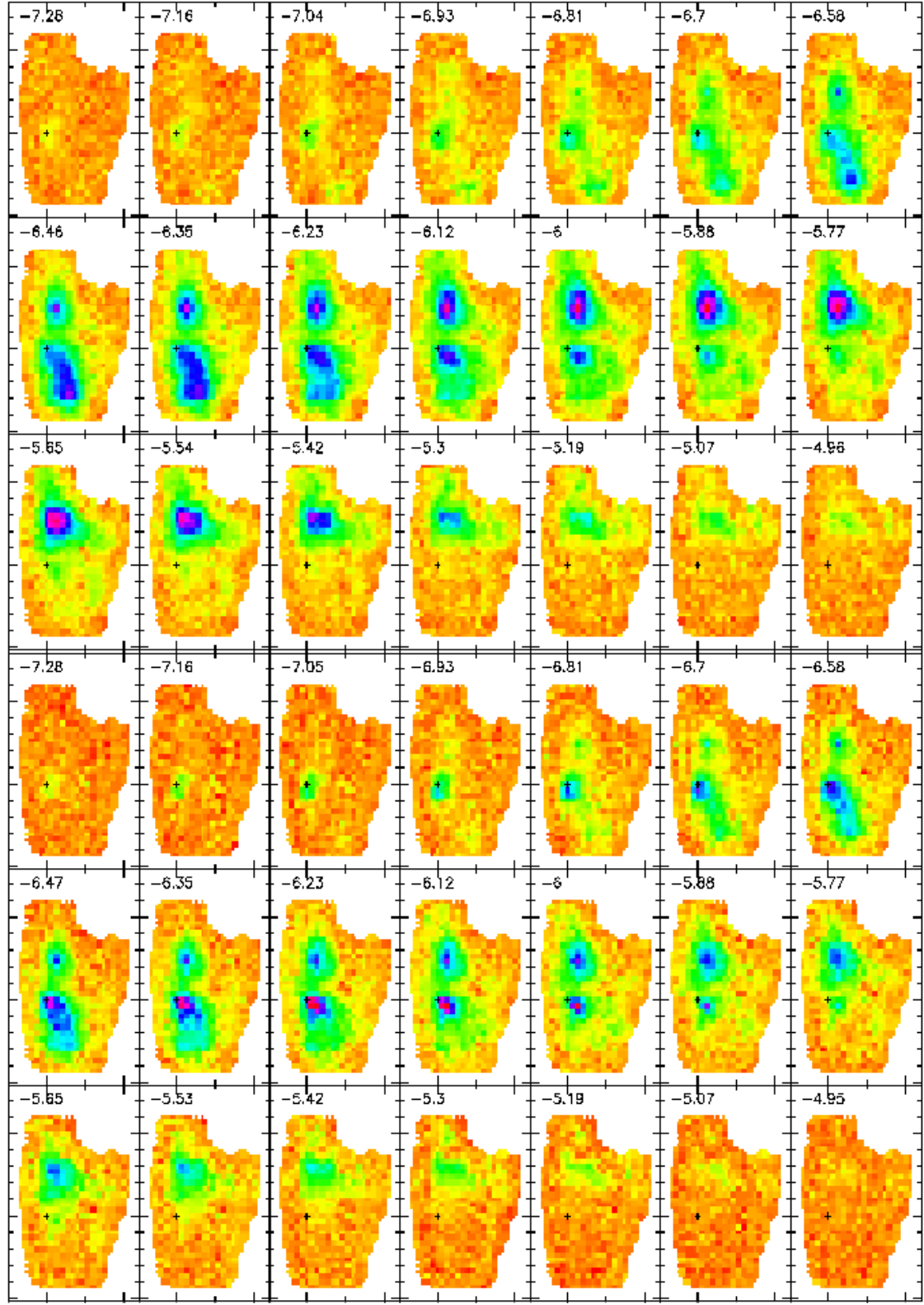} \caption{\tastar channel
   map of the C$^{18}$O~(1--0) emission~(upper three rows) and
   C$^{18}$O~(2--1)~(lower three rows). The colour scale is the same for both
   transitions.  The LSR velocity is indicated
   in the upper left corners of the panels. Each pixel corresponds to
   a single observed position but the C$^{18}$O~(2--1) data is binned
   in this figure to the same channel width as the (1--0)
   transition, 0.116 \kmps.  The highest intensity in the panels is
   $\sim3.0$~K.  The cross in the panels is located in the map centre
   position.}
   \label{figure:c18o_1-0_channel} 
   \end{figure*}

CG12-SW is inconspicuous when compared to the two
stronger \ceo maxima.  No signs of star formation have
been found in its direction and it is therefore either still in
pre-star formation phase or its density-temperature structure is such
that no star formation will take place.  The arc like feature which connects
\cgs and \cgsw at velocities from $-6.6$ \kmps\ to $-6.2$ \kmps \ in
Fig.  \ref{figure:c18o_1-0_channel} suggests that \cgsw might be connected
with  \cgsp.

% \subsubsection{CG 12 in comparison to other star forming regions}

%At the distance of  630~pc to CG 12 
%(Williams et al. 1977) the $6\arcmin$ by
%10$\arcmin$ apparent size of the \ceo cloud translates to 1.1~pc by
%1.8~pc.  The \ceo observations cover only
%the cloud compact head but the cloud extends further to the South
%(\cite{yonekuraetal1999a}).  If  CG 12 were at the distance of 150~pc 
%the \thco cloud (Fig. \ref{figure:CG12_13CO_map}) would
%appear to be as large as the low mass star formation regions  
%Chamaeleon I cloud south of the Ced 110 
%reflection nebulosity (e.g. Haikala et al. 2005) or CrA
%(e.g. \cite{harjuetal1993}). The size of CG 12 in \twco including the
%tail as observed by \cite{yonekuraetal1999a} is 14$\arcmin$ by 40
%$\arcmin$.  The CG 12 mass is likely to
%be larger than 100 M$_{\sun}$.  The binary h4636 consisting of a B7 and a
%B4 star together with several lower mass stars (Williams et al. 1977)
%have already formed in the region. Further star formation is still
%going on as indicated by the two associated IRAS point sources, the
%two compact 1.2 mm continuum sources and by the detected molecular
%outflow(s).

% Like CrA, CG 12 presents a cometary structure, an elongated head
% where star formation is taking place and a $\sim 1 \degr$ long
% tail. CrA also harbors a large molecular outflow and 
%numerous mm continuum sources (e.g. \cite{levreault1988},
% \cite{chinietal2003}).

\section{Conclusions}\label{sec:conclusions}

We have performed a detailed, high signal-to-noise ratio, mm line study
of CG 12 in various molecular transitions, principally of
C$^{18}$O~(1--0) and C$^{18}$O~(2--1), as well as in molecular
lines probing dense material, and have obtained the following results:

   1. The C$^{18}$O line emission is distributed in a 10$\arcmin$ North-South
   elongated lane with two strong, compact maxima, \cgn and
   \cgsp, and a weaker maximum, \cgswp.

   2. High density tracers CS (2--1), (3-2), \dcopl and \htcopl are
   detected in both strong \ceo maxima. Emission from these molecules
   is weak in \cgn but in \cgs it defines a compact core (referred to
   as \dcopl core) which is spatially offset from the \ceo maximum.

   3. The emission from the high density tracers in \cgn takes place
   at a velocity where emission from \ceo is weak. The molecules
   associated with the cloud component  detected in high density tracers are
   likely to be heavily depleted.  

   4. Positive Matrix Factorization was applied to study the cloud
      \ceo fine scale structure. The observed strong \ceo emission in
      \cgn (PMF factor 2{\it n}) originates in the envelope of the
      depleted cloud component or in a separate entity seen in the
      same line of sight.  In \cgs the most intense PMF factor 3{\it
      s} traces warm gas on the surface of the \dcopl core or a
      separate adjacent cloud component.

   5. The driving source of the collimated molecular outflow detected
   by \cite{white1993} lies in the \dcopl core.

   6. The average \ceo LTE mass is $\sim$80 M$_{\sun}$ for \cgnp,
   $\sim$70 M$_{\sun}$ for \cgs and $\sim$20 M$_{\sun}$ for
   \cgswp. These numbers can only be considered as a zeroth order
   estimate because of the uncertainty in defining the \ceo excitation
   temperature and possible molecular depletion in the \ceo maxima. 

   7. If the distance to CG 12, 630 pc, is correct the linear size and
   the mass of this cometary globule approaches that of a typical
   low mass star forming region like eg. Chamaeleon I.

\begin{acknowledgements} 

 We thank Bo Reipurth and the A\&A Letters editor, Malcolm Walmsley for
 critically reading the manuscript and for very useful comments that
 helped to improve this paper.  We also thank Mika Juvela and Kalevi
 Mattila for helpful discussions and Pentti Paatero for providing the
 PMF code to us. Data retrieved from the Canadian Astronomy Data
 Centre were used to produce the CO outflow contours in
 Fig. \ref{figure:sofi}. Canadian Astronomy Data Centre is operated by
 the Herzberg Institute of Astrophysics, National Research Council of
 Canada.

\end{acknowledgements}

\listofobjects
\Online

\listofobjects

\appendix

\section {Positive Matrix Factorization analysis of CG 12} \label{sec:PMF_app}

\subsection{\cgn}

\begin{figure} \centering \includegraphics 
%[width=8cm]{cgn_PMF_10_2}
[width=8cm]{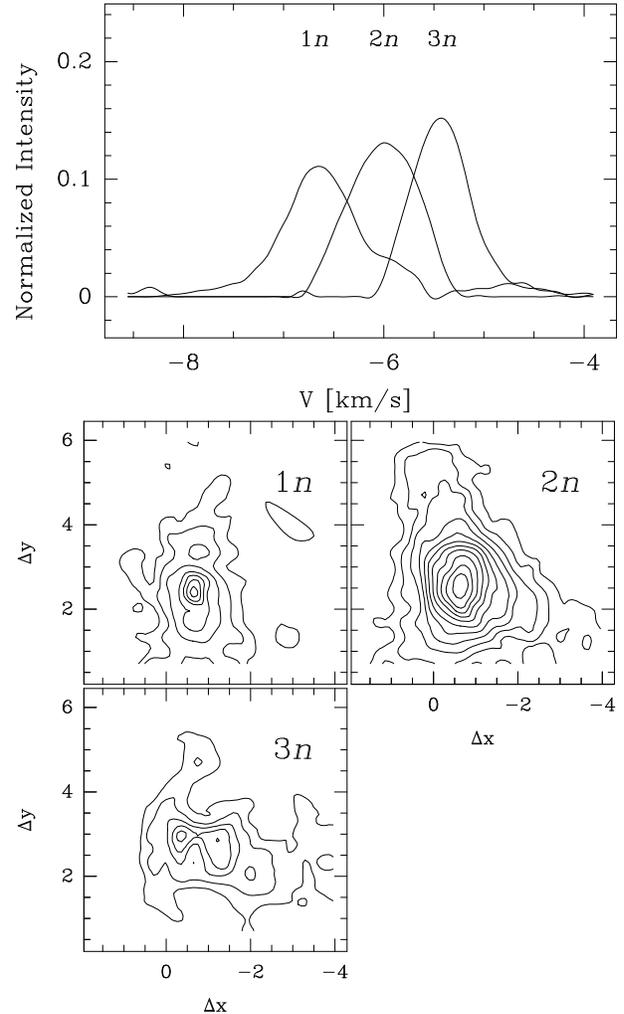}
\caption{ Upper panel: the basic spectral profiles (factors)
calculated by the PMF in \cgn~(three factor fit), \ceo(1--0)
data. Lower panels: maps of the weights of the basic factors. The
lowest contour level and increment are 0.2 \kkms.  Offsets from the
centre position of the original \ceo map are shown in arcminutes on
the x and y axes.}

\label{figure:cgn_PMF_10}
\end{figure}

\begin{figure} \centering \includegraphics 
%[width=8cm]{cgn_PMF_21_2}
[width=8cm]{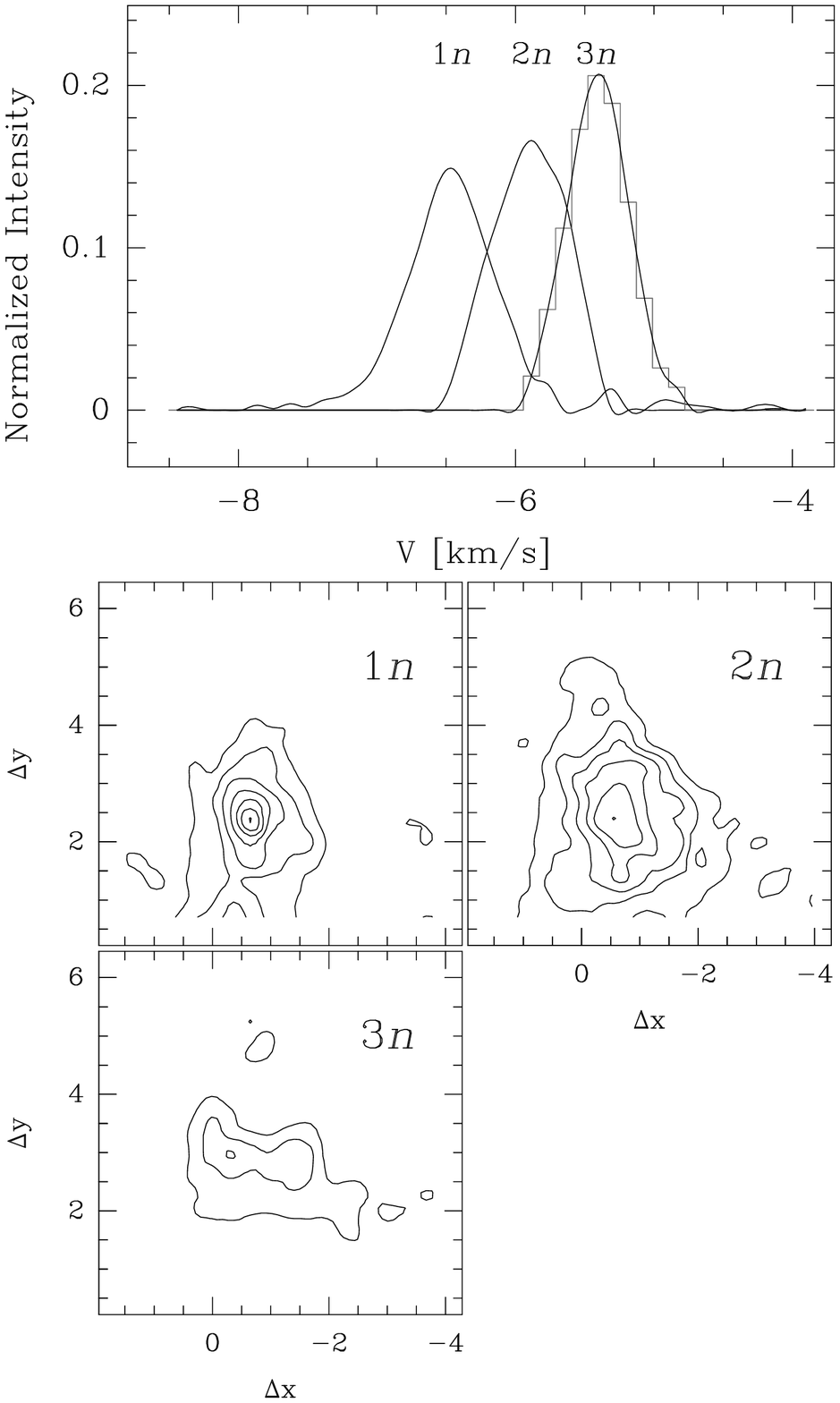}
\caption{ As Fig.~\ref {figure:cgn_PMF_10} but for \ceo(2--1) data.
Factor 3{\it n} in the upper panel has been plotted also as a
histogram to indicate velocity resolution of the input/output spectra.
}
\label{figure:cgn_PMF_21}
\end{figure}

 The results of the PMF, assuming three factors, to observed
\ceo~(1--0) and~(2--1) data in \cgn are shown in Figs. 
\ref{figure:cgn_PMF_10} and \ref{figure:cgn_PMF_21}. The uppermost
panels show the three fitted PMF factors.  The area under each factor
is normalized to one.  The histogram superposed on the rightmost
factor in Fig~\ref{figure:cgn_PMF_21} indicates the velocity resolution
of the input data. The intensity distribution of the PMF factors in
\cgn are shown in the lower three panels.

 The PMF reproduces well the velocity structure displayed in
Fig. \ref{figure:c18o_1-0_channel}. The fits to the \ceo ~(1--0) and
(2--1) data are independent of each other. Therefore it is encouraging
that the PMF output factors for both \ceo transitions are similar in 
velocity and in spatial distribution.  Because of the unambiguity
of the fits for both transitions the factors, centered at $-6.6$, $-5.9$
and $-5.4$ \kmps, are referred to in the following as {1\it n}, 2{\it n} and 
3{\it n},
respectively.
%**********************************************************************
%**********************************************************************
%**********************************************************************


 The PMF factors, 1{\it n}, 2{\it n} and 3{\it n}, are distributed in a
north-south oriented ridge, in a pear shaped body and a narrow east-west
ridge, respectively. The 3{\it n} factor lies at the velocity where the
emission from the \htcoplp, \dcoplp, and \nthpl is at maximum.  Even
though this factor is not readily evident in the individual spectra
PMF finds it in both \ceo transitions.  Factors 2{\it n} and 3{\it n} are
symmetric but factor 1{\it n} has blue and redshifted wing like
emission, especially in the~(1--0) transition. According to the PMF fit 
the high \tastar(C$^{18}$O~(1--0))/\tastar(C$^{18}$O~(2--1)) 
ratio observed in the direction of \cgn is due to the factor
2{\it n}.

 If the number of factors in the PMF is chosen to be four PMF divides in
essence the factor 1{\it n} into two, leaving the two other factors
untouched. The spatial distribution of the split up factors is similar
to that of the 1{\it n} in the three-factor PMF. If the number of factors is
further increased to five the result is similar to that with four
factors plus a fifth factor which has nearly zero intensity, 'empty'
field, in the map.  The three factors are therefore sufficient to
produce the velocity structure evident in
Fig.~\ref{figure:c18o_1-0_channel} and increasing the number of
factors does not improve the fit significantly.

\subsection{\cgs and \cgsw}

\begin{figure} \centering \includegraphics 
[width=8cm]{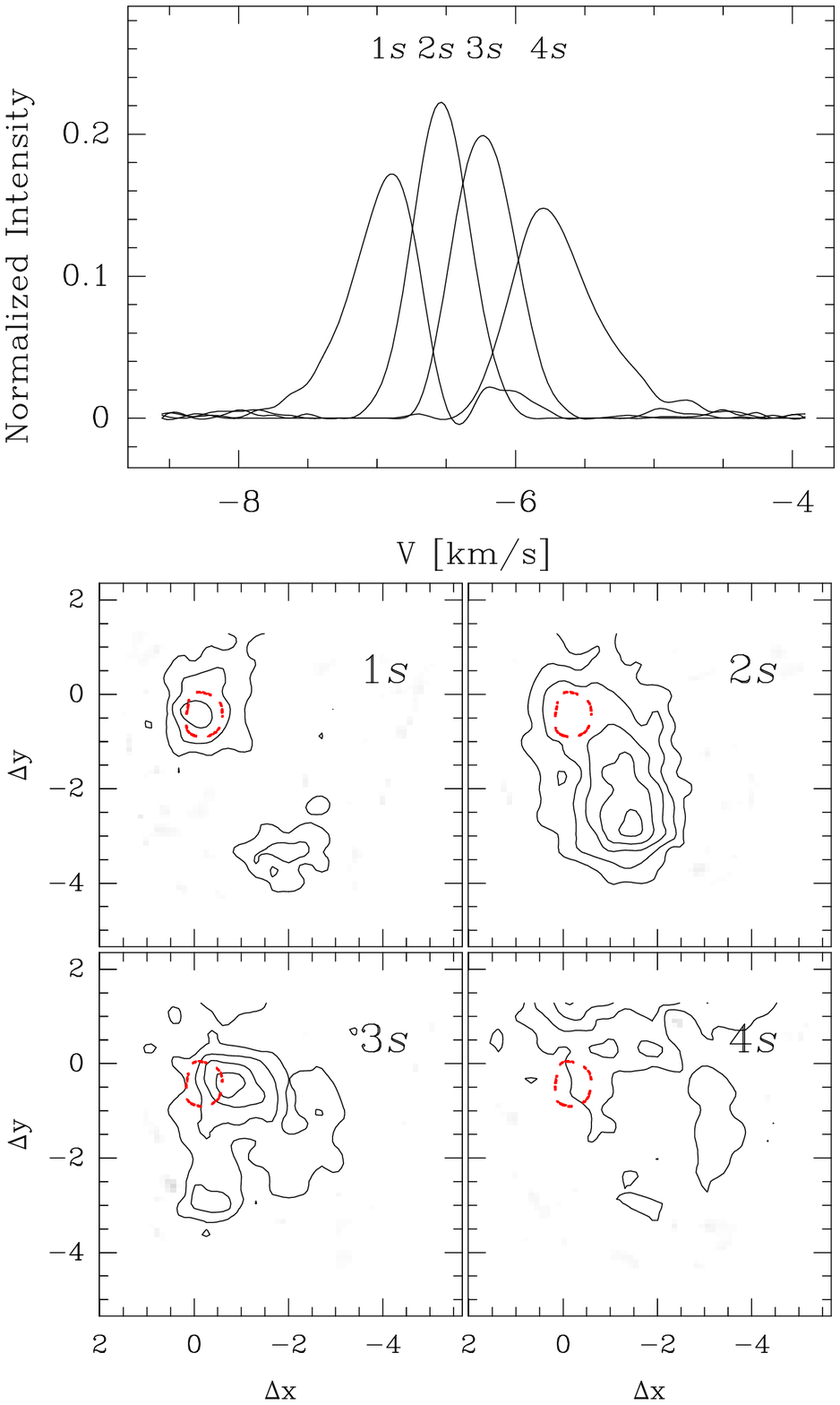}
\caption{ Upper panel: the basic spectral profiles calculated by the
PMF in \cgs~(four factor fit), \ceo(1--0) data. Lower panels: the maps
of the intensity of the basic factors. The factor numbers correspond
to the numbers in the upper panel.  The lowest contour level and
increment is 0.2 \kkms. The dashed contour outlines the \dcopl core.
}
\label{figure:cgs_PMF_10}
\end{figure}

\begin{figure} \centering \includegraphics 
[width=8cm]{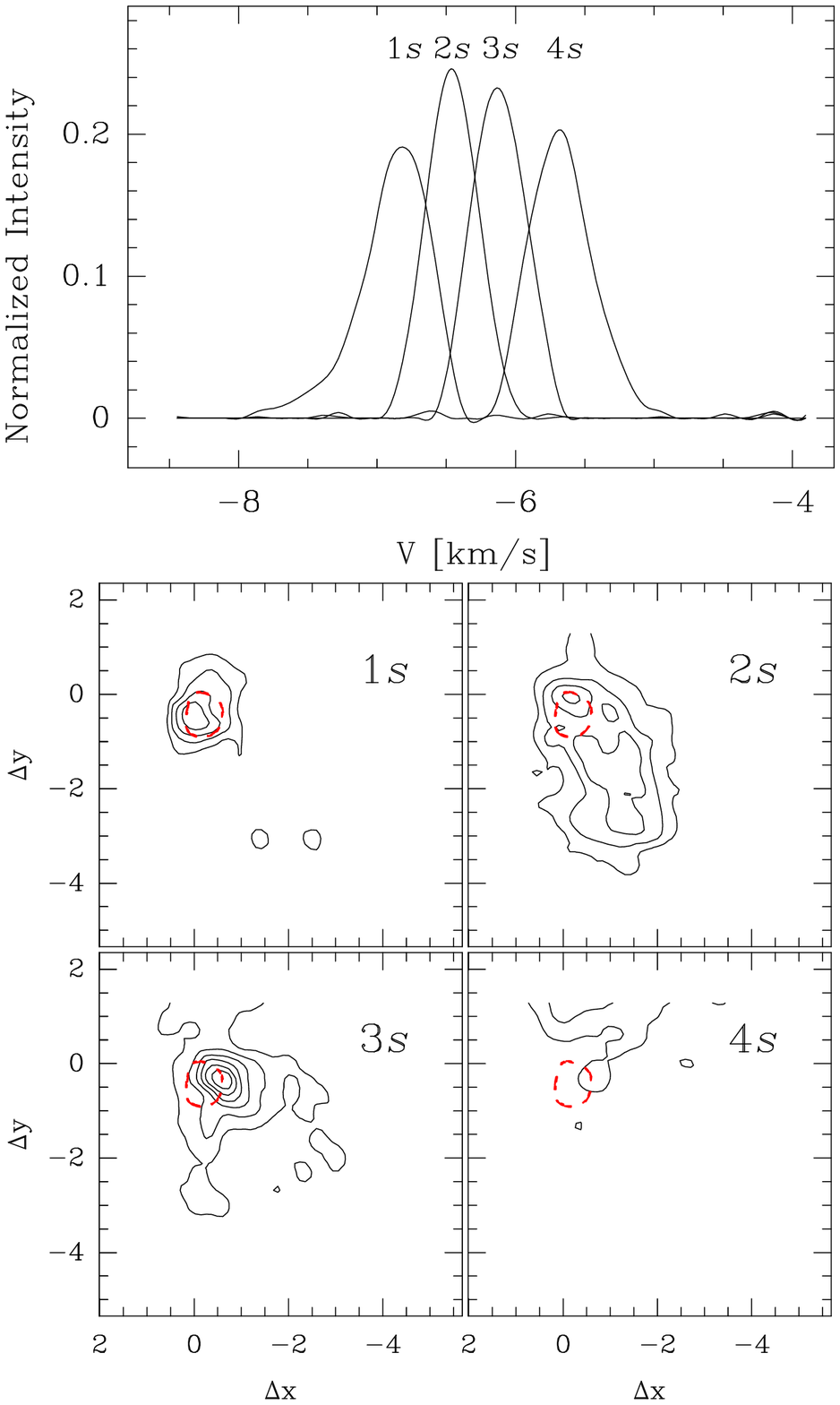}
\caption{As Fig.~\ref {figure:cgs_PMF_10} but for \ceo(2--1) data.
}
\label{figure:cgs_PMF_21}
\end{figure}

 The results of the PMF fits, assuming four factors, to both observed
\ceo transitions in \cgs are shown in Figs.~\ref{figure:cgs_PMF_10}
and \ref{figure:cgs_PMF_21}.  The location of the \dcopl core is shown
with a dashed contour in the figures. The factors, centered at $-6.8$,
$-6.4$, $-6.2$ and $-5.8$ \kmps, will be referred to as 1{\it s} to
4{\it s}, respectively.

 The most redshifted factor 4{\it s} covers the very northern part of the
region and is close in velocity to the 2{\it n} factor in \cgnp. It is
natural to consider that this factor is due to emission extending from
\cgn to \cgsp.  Also the very faint emission observed in the northern
part of the field in other three factors is at least partly due to
emission from \cgnp.  

The emission from the most blueshifted factor 1{\it s} is concentrated
just below the (0,0) position.  The 2{\it s} factor represents the
arc seen in Fig.~\ref{figure:c18o_1-0_channel} which connects \cgs and
\cgsw.  There is a small size local maximum in the \ceo~(2--1) 2{\it
s} factor north of the \dcopl core. Even though this local maximum is
not seen in the \ceo~(1--0) 2{\it s} factor, there is extended
emission at its location.  The southern part of the arc is more
intense in the (1--0) transition than in the (2--1). The maximum of
the southern extension coincides with \cgswp.  The emission from the
3{\it s} factor peaks west of the \dcopl core and is seen in both
transitions.
% Further, this maximum lies in the local \ceo~(2--1) dip
%between the maximum and southern extension of the 2{\it s} \ceo~(2--1)
%factor.

\begin{figure*} \centering \includegraphics [ bb= 80 20 550 780,
width=12.5cm, angle=-90, clip]{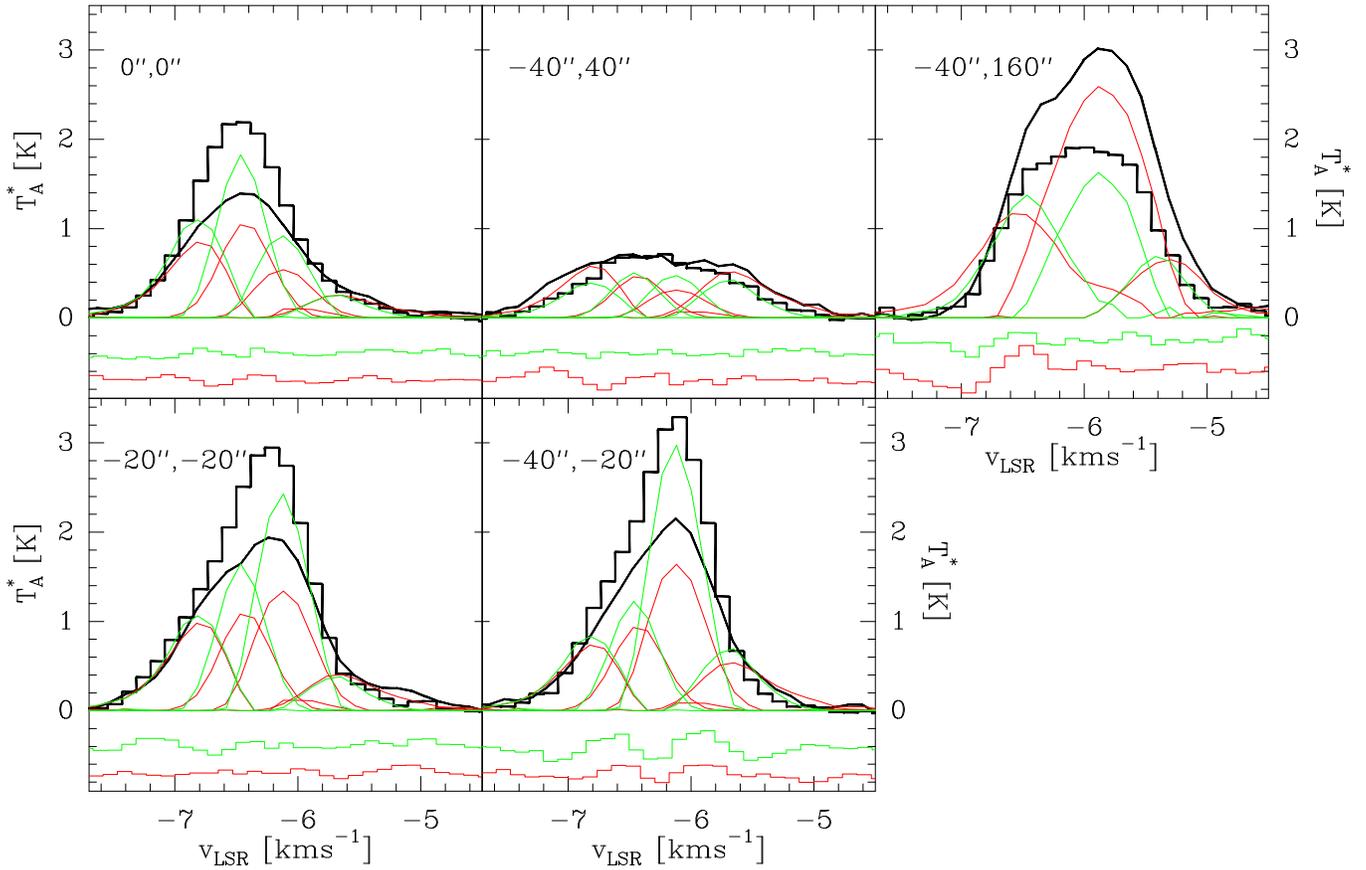}
\caption{ The PMF fit to the \ceo lines in the selected positions
positions in Fig.~\ref{figure:spectra2}. The observed \ceo (2--1) and
(1--0) lines are plotted with a heavy histogram and line, respectively.
The green  and  red lines show the fitted PMF \ceo (2--1) and (1--0) factors,
respectively. The residuals of the fits are shown in the
lower part of each panel.}
\label{figure:PMFspectra}
\end{figure*}

\begin{figure} \centering \includegraphics [ bb= 80 130 550 500,
width=9.5cm, clip]{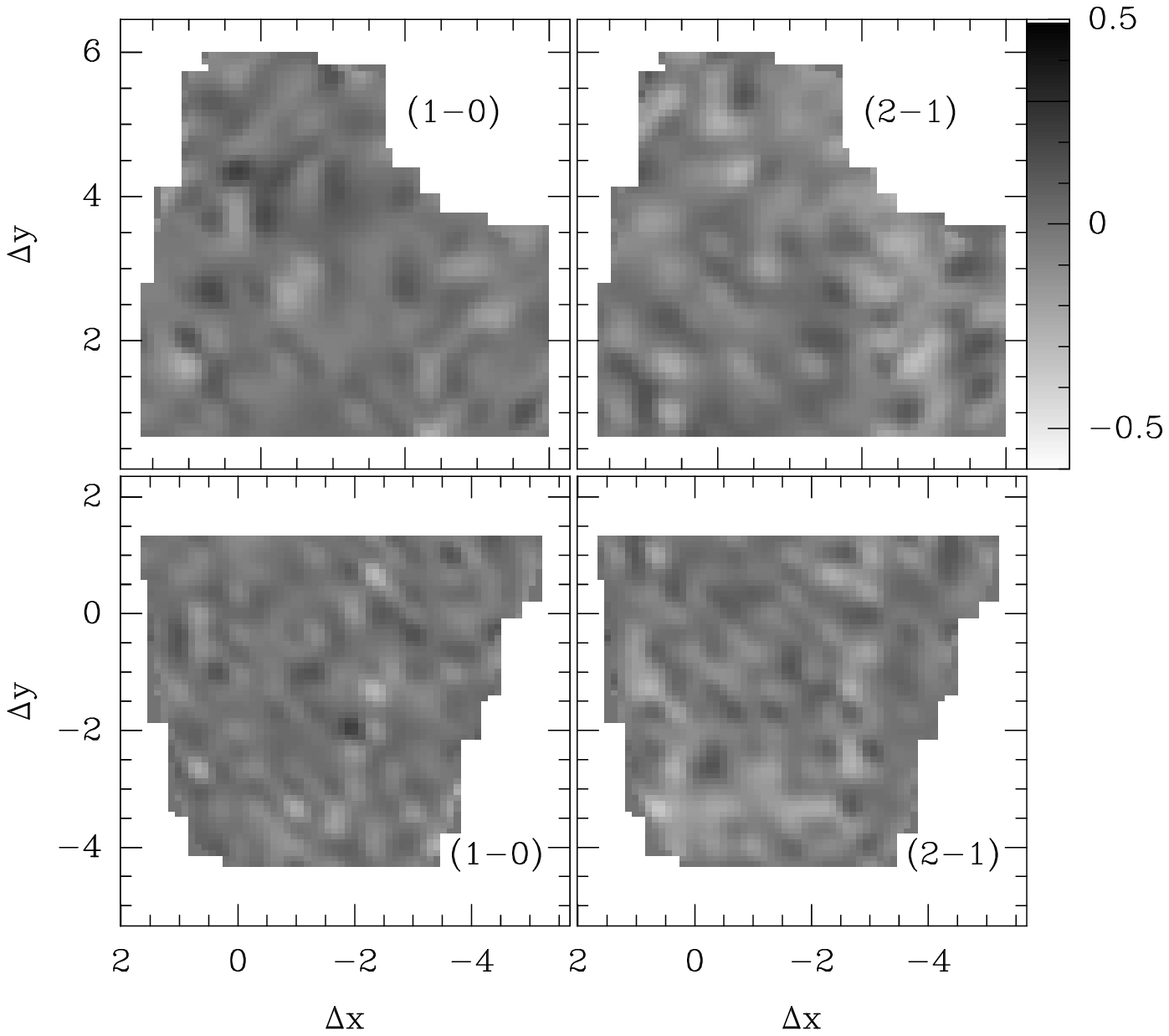}
\caption{ The PMF fit residuals in the \kkms scale.}
\label{figure:PMFresiduals}
\end{figure}

\subsection{Individual line profiles}

 The PMF fits to the \ceo lines in the same selected positions as  in
Fig.~\ref{figure:spectra2} are shown in Fig.~\ref{figure:PMFspectra}.  The
residuals obtained by subtracting the added up PMF factors from the
observed lines are  shown below the lines in the figure. A grey scale
map of all the PMF fit residuals is shown in
Fig.~\ref{figure:PMFresiduals}. 
One should stress that, unlike a
gaussian multicomponent fit to a single spectrum, PMF fits, in this
case 3~(\cgnp) or 4~(\cgsp) factors into all the input spectra
simultaneously. Neither the shape nor the velocity of the factors is
fixed in the fit. The two \ceo transitions were fit separately and PMF
could have used factors differing in profile and velocity for the two
transitions.  The fitted PMF factors are, however, similar both in shape and
velocity.  The residuals shown in Figs.~\ref{figure:PMFspectra}
and \ref{figure:PMFresiduals} are small and demonstrate that PMF
produces a good fit to the data.

The largest residual in Fig.~\ref{figure:PMFspectra} takes place in
the blue shifted side of the  \ceo~(1--0) line in the centre of \cgn~(position
$-$40\arcsec,160\arcsec). This would imply that the blue shifted wing
of factor \ceo (1--0) 1{\it n} is too strong. A closer comparison of
the input data and PMF results reveals that similar deviations take
place for three other \ceo~(1--0) spectra around the position
above. The \ceo~(1--0) line profiles farther away from the \cgn centre
actually do have a more prominent blue shifted wing than the four
above mentioned spectra. The large number of spectra with good fits
outweighs the less optimal fit for the spectra in the centre of
\cgnp. The residuals shown in Fig.~\ref{figure:PMFspectra} for the remaining
four spectra in \cgs are smaller than in
position~($-$40\arcsec,160\arcsec).

% It would have been possible to fit each of the observed lines using
%two or three gaussian profiles. 
%However, if all the gaussian
%parameters are allowed to vary  unless some of the fitted parameters is
%fixed
%The situation
%would be the same had the lines been fit with gaussian profiles

\end{document}